\documentclass[fleqn]{2023SCGE}
\makeatletter
\def\@addressline#1#2{\@addresscr\llap{\textsuperscript{\rm#1}\kern0.25em}#2}
\def\@addresscr{\def\\{\tabularnewline\@addresscr\ignorespaces}}
\makeatother

\begin{document}

\ensubject{Astronomy and Astrophysics}
\ArticleType{Article}
\SpecialTopic{}
\Year{XXX}
\Month{XXX}
\Vol{XXX}
\No{XXX}
\DOI{XXX}
\ArtNo{XXXXXX}
\ReceiveDate{19-Oct-2026}
\AcceptDate{16-Nov-2026}

\title{LAMOST III Low-resolution Spectroscopic Survey: Overview and Target Selection}{LAMOST III Low-resolution Spectroscopic Survey: Overview and Target Selection}

\author[1,2]{Shuai Xu}{}
\author[1,2]{Haibo Yuan}{yuanhb@bnu.edu.cn}
\author[3,1]{Maosheng Xiang}{}
\author[4]{Bingqiu Chen}{}
\author[5]{Yang Huang}{}
\author[6]{Chun Wang}{}
\author[3]{\\Hailong Yuan}{}
\author[7,2]{Tianqi Cang}{}
\author[8,5,9]{Xuefei Chen}{}
\author[3,1,5]{Yuqin Chen}{}
\author[10]{Yao-Quan Chu}{}
\author[11]{Wenyuan Cui}{}
\author[12,13, 3]{\\Subo Dong}{}
\author[14,15]{Min Fang}{}
\author[16]{Shuai Feng}{}
\author[1,2]{Jianning Fu}{}
\author[17]{Wei-Min Gu}{}
\author[3]{Yan-Xin Guo}{}
\author[8,9]{\\Zhanwen Han}{}
\author[18,5]{Jinliang Hou}{}
\author[3]{Wen Hou}{}
\author[1,2,3]{Bowen Huang}{}
\author[3]{Zhiying Huo}{}
\author[3]{Xiao Kong}{}
\author[15]{\\Xu Kong}{}
\author[3]{Guangwei Li}{}
\author[3]{Haining Li}{}
\author[8,9]{Jiangdan Li}{}
\author[8,9]{Jiao Li}{}
\author[3]{Chao Liu}{}
\author[3,5,19]{Jifeng Liu}{}
\author[20]{\\Qichun Liu}{}
\author[4]{Xiao-Wei Liu}{}
\author[1,2]{Xue Lu}{}
\author[3]{Ali Luo}{}
\author[5,3]{Zexi Niu}{}
\author[13]{Dan Qiu}{}
\author[18]{Shiyin Shen}{}
\author[3]{Jianrong Shi}{}
\author[21]{\\Botian Tang}{}
\author[3,1]{Hao Tian}{}
\author[22,23]{Haijun Tian}{}
\author[24]{Zhijia Tian}{}
\author[25,3,5]{Jian-Min Wang}{}
\author[3,5]{Jie Wang}{}
\author[20]{\\Xiaofeng Wang}{}
\author[1,2]{Yuanchang Wang}{}
\author[3]{Hong Wu}{}
\author[2,1]{Jianghua Wu}{}
\author[12,13]{Xue-Bing Wu}{}
\author[20]{Qiqi Xia}{}
\author[5,3]{\\Kai Xiao}{}
\author[26,27]{Jiwei Xie}{}
\author[3,1,5]{Hongliang Yan}{}
\author[28]{Lin Yang}{}
\author[29,30]{Xiaohu Yang}{}
\author[3]{Haotong Zhang}{}
\author[12,13]{\\Huawei Zhang}{}
\author[18]{Hui Zhang}{}
\author[31,32]{Liyun Zhang}{}
\author[3]{Meng Zhang}{}
\author[33]{Zhi-Xiang Zhang}{}
\author[3,5]{Gang Zhao}{}
\author[3,5,1]{\\Gong-Bo Zhao}{}
\author[34,35]{He Zhao}{}
\author[3]{Yong-Heng Zhao}{}
\author[18]{Zhenya Zheng}{}
\author[18]{Jing Zhong}{}
\author[1,2]{Weikai Zong}{}

\AuthorMark{Xu S}
\AuthorCitation{Xu S, Yuan H B, Xiang M S, et al.}

\address[1]{Institute for Frontiers in Astronomy and Astrophysics, Beijing Normal University, Beijing 102206, China}
\address[2]{School of Physics and Astronomy, Beijing Normal University, Beijing 100875, China}
\address[3]{National Astronomical Observatories, Chinese Academy of Sciences, Beijing 100101, China}
\address[4]{South-Western Institute for Astronomy Research, Yunnan University, Kunming 650500, China}
\address[5]{School of Astronomy and Space Science, University of Chinese Academy of Sciences, Beijing 100049, China}
\address[6]{Tianjin Astrophysics Center, Tianjin Normal University, Tianjin 300387, China}
\address[7]{Department of Scientific Research, Beijing Planetarium, No.138 Xizhimenwai Street, 100044, Beijing, China}
\address[8]{Yunnan Observatories, Chinese Academy of Sciences, Kunming 650216, China}
\address[9]{International Centre of Supernovae (ICESUN), Yunnan Key Laboratory, Kunming 650216, China}
\address[10]{University of Science and Technology of China, Hefei 230026, China}
\address[11]{School of Physics, Hebei Normal University, Shijiazhuang 050024, China}
\address[12]{Department of Astronomy, School of Physics, Peking University, Beijing 100871, China}
\address[13]{Kavli Institute for Astronomy and Astrophysics, Peking University, Beijing 100871, China}
\address[14]{Purple Mountain Observatory, Chinese Academy of Sciences, Nanjing 210023, China}
\address[15]{School of Astronomy and Space Science, University of Science and Technology of China, Hefei 230026, China}
\address[16]{College of Physics, Hebei Normal University, Shijiazhuang 050024, China}
\address[17]{Department of Astronomy, Xiamen University, Xiamen, Fujian 361005, China}
\address[18]{Shanghai Astronomical Observatory, Chinese Academy of Sciences, Shanghai 200030, China}
\address[19]{New Cornerstone Science Laboratory, National Astronomical Observatories,\\ Chinese Academy of Sciences, Beijing 100101, China}
\address[20]{Physics Department, Tsinghua University, Beijing 100084, China}
\address[21]{School of Physics and Astronomy, Sun Yat-sen University, Zhuhai 519082, China}
\address[22]{School of physics, Hangzhou Dianzi University, Hangzhou 310018, China}
\address[23]{Zhejiang Branch of National Astronomical Data Center, Hangzhou 310018, China}
\address[24]{Department of Astronomy, Key Laboratory of Astroparticle Physics of Yunnan Province,\\ Yunnan University, Kunming 650200, China}
\address[25]{Key Laboratory for Particle Astrophysics, Institute of High Energy Physics,\\ Chinese Academy of Sciences, Beijing 100049, China}
\address[26]{School of Astronomy and Space Science, Nanjing University, Nanjing 210023, China}
\address[27]{Key Laboratory of Modern Astronomy and Astrophysics, Ministry of Education, Nanjing 210023, China}
\address[28]{Department of Cyber Security, Beijing Electronic Science and Technology Institute, Beijing, 100070, China}
\address[29]{State Key Laboratory of Dark Matter Physics, Tsung-Dao Lee Institute \& School of Physics and Astronomy,\\ Shanghai Jiao Tong University, Shanghai 201210, China}
\address[30]{Shanghai Key Laboratory for Particle Physics and Cosmology, and Key Laboratory for Particle Physics,\\ Astrophysics and Cosmology, Ministry of Education, Shanghai Jiao Tong University, Shanghai 200240, China}
\address[31]{College of Physics, Guizhou University, Guiyang 550025, China}
\address[32]{Guizhou Radio Astronomical Observatory, Guizhou University, Guiyang 550025, China}
\address[33]{College of Physics and Information Engineering, Quanzhou Normal University, Quanzhou 362000, China}
\address[34]{Institute of Astronomy and Physics, Inner Mongolia University, Hohhot 010021, China}
\address[35]{Departamento de Fisica y Astronomia, Facultad de Ciencias Exactas, Universidad Andres Bello,\\ Fernandez Concha 700, 7591538 Las Condes, Santiago, Chile}

\abstract{
We present an overview of the LAMOST~III Low-resolution Spectroscopic Survey (LRS), including its scientific goals, target-selection strategy, and current observing status.
Targets in the main survey are selected via weighted random sampling, using the inverse local density in the Hertzsprung--Russell (HR) diagram and in the $G$-magnitude distribution, with additional enhancement in selected regions of the HR diagram.
The main survey is complemented by 35 dedicated add-on catalogs (including very metal-poor star candidates, halo stars, hot/young stars, wide binaries, extragalactic sources and so on) and a time-domain program providing multi-epoch spectroscopy of binaries and other peculiar systems.
LAMOST~III LRS is designed to map the structure and evolution of the Galactic disk, reconstruct the assembly and early chemical-enrichment history of the Galactic halo, improve stellar-parameter calibration, and characterize rare and variable objects.
By June 2025, the LAMOST~III LRS had obtained 1.53 million low-resolution spectra from 658 plates, corresponding to 1,264,052 unique sources.
Relative to LAMOST~I+II, the current sample is enriched in O/B stars, subgiants, main-sequence turn-off stars, metal-poor stars, and other previously underrepresented populations.
In particular, the fractions of metal-poor and very metal-poor stars increase from 3.8\% and 0.4\% in LAMOST~I+II to 9.0\% and 1.1\% in LAMOST~III LRS, respectively, demonstrating the improved efficiency of the revised target-selection strategy for Galactic archaeology and stellar astrophysics.}


\thispagestyle{empty}
\begingroup
\makeatletter
\long\def\@makefntext#1{\noindent#1}
\footnotetext[0]{\qihao *Corresponding author (email:~\@authoremail)}
\makeatother
\endgroup
\maketitle

\begin{multicols}{2}
\section{Introduction}\label{sec:intro}
The Large Sky Area Multi-Object Fiber Spectroscopic Telescope (LAMOST; \cite{cui2012,deng2012,Zhao2012,liu2014}) is a 4\,m quasi-meridian reflective Schmidt telescope with 4000 fibers feeding a 5$^{\circ}$ field-of-view focal plane.
Since commissioning, LAMOST has operated through the pilot survey (2011.10.24--2012.06.17), LAMOST~I (2012.09.28--2017.06.16), and LAMOST~II (2017.09.19--2023.06.15), collecting more than 20 million low- and medium-resolution stellar spectra between October~2011 and September~2024.
The eleventh data release (DR11)\footnote{\url{https://www.lamost.org/dr11/v2.0/}} provides 11.93 million low-resolution spectra from 6,283 observing fields and 13.16 million medium-resolution spectra from 2,395 fields.
For the low-resolution survey (LRS), DR11 also includes stellar-parameter estimates for approximately 11.57 million sources.

With a dataset now exceeding several tens of millions of spectra, LAMOST is among the most productive spectroscopic survey facilities worldwide.
As reviewed by Yan et al.~\cite{yan2022}, LAMOST observations over the past decade have driven advances across a broad range of astrophysical topics, including the structure, formation, and evolution of the Milky Way (e.g.\ \cite{gallart2019,xiang2022,jiadon2023,huang2024a,xiang2025,chen2025,2025zhang,he2026}), stellar astrophysics and chemical evolution (e.g.\ \cite{yan2018,xing2019,yan2021,xing2023,chen2023}), exoplanetary systems (e.g.\ \cite{tu2025,chen2026,2026shin}), and the discovery of rare and compact stellar objects (e.g.\ \cite{liu2019,Rebassa2019,huang2021,yi2022,yaguang2022,martin2022,wang2024,hern2025,huang2025nsr,yu2025}).

Early LAMOST target selection adopted uniform sampling in the color--magnitude diagram \cite{yuan2015}, constrained by the photometric surveys available at the time---primarily the Sloan Digital Sky Survey (SDSS; \cite{SDSS2000,Yanny2009}), the 2MASS All-Sky Point Source Catalog (2MASS; \cite{tmass2006}), Pan-STARRS1 (PS1; \cite{ps1}), and the Xuyi Schmidt Telescope Photometric Survey of the Galactic Anti-center (XSTPS-GAC; \cite{liuxw2014,zhang2014}).
Although effective for constructing large samples with a relatively simple and well-defined selection function \cite{Chen2018}, this strategy underrepresented several key populations. 
Metal-poor (MP; [Fe/H] $<-1$) and very metal-poor (VMP; [Fe/H] $<-2$) stars constitute only $\sim$3.8\% and 0.4\% of the sample, respectively, with the VMP stars corresponding to approximately $2\times10^4$ in total. 
Giant stars account for $<20$\%, and distant stars are underrepresented: half of the sample lies within 1~kpc, and only 1.5\% lies beyond 10~kpc, implying comparatively low fractions of outer-disk and halo stars. Young stellar objects and OB-type stars are likewise scarce, comprising $\sim$0.15\% and $\sim$4\%, respectively.
In addition, the relatively low signal-to-noise ratio (S/N) of many LAMOST spectra limits the precision of derived stellar ages and chemical abundances. For example, only 24\% of the DR11 spectra reach $\mathrm{S/N}_g>50$, mainly for stars brighter than $G\sim15$.

In the past decade, complementary surveys---including Gaia \cite{gaiadr1}, the Stellar Abundance and Galactic Evolution Survey (SAGES; \cite{fan2023}), the SkyMapper Southern Survey (SMSS; \cite{skymapper}), the Javalambre Local Universe Photometric Survey (J-PLUS; \cite{jplus}), and GALEX---have delivered precise distances, proper motions, and multi-band photometry for hundreds of millions of stars.
These data products enable the identification of large numbers of peculiar objects and the inference of stellar parameters; for example, Gaia multi-band photometry and slitless spectra provide metallicities and other parameters for nearly 100 million stars \cite{huang2022,xu2022,yang2022,huang2023,Zhanggreen2023,huang2024,lu2024,huang2025,yang2025,sun2025}.
The availability of these rich datasets enables more accurate and efficient target selection for LAMOST~III LRS.

Building on these developments, LAMOST~III LRS officially commenced in September 2023, with its completion date yet to be determined. Leveraging the new generation of photometric and astrometric catalogs, it expands coverage of metal-poor stars, giants, halo stars, main-sequence turn-off and subgiant stars, young stars, cluster members, wide binaries, high-tangential-velocity stars, and anomalous objects in the Hertzsprung--Russell diagram. Relative to LAMOST~I/II, the corresponding sample sizes increase by factors of several to tens, improving completeness and sampling efficiency and enhancing the survey's discovery potential in Galactic structure and evolution, stellar astrophysics, and rare-object searches.

In this paper, we present the scientific goals and target-selection strategy of LAMOST~III LRS, together with a brief assessment of observations obtained up to June~30, 2025. Section~\ref{sec:goals} outlines the scientific objectives. Section~\ref{subsec:HRD} describes the input stellar catalog, the selection strategy, and results for the main survey. Section~\ref{sec:add_on} presents the add-on catalogs, their simulated observations, and the time-domain (TD) survey. Section~\ref{sec:now} summarizes the current observational status, and Section~\ref{sec:summary} summarizes the survey and discusses future prospects.

\section{Scientific Goals}\label{sec:goals}

LAMOST~III LRS is designed to leverage large-scale spectroscopy together with the astrometric and photometric information provided by Gaia and complementary surveys.
By targeting distant tracers, rare stellar populations, and other high-value sources that were underrepresented in LAMOST~I/II, the survey aims to address two core Galactic-science objectives: (1) mapping the multi-dimensional structure and evolution of the Galactic disk, and (2) reconstructing the assembly and early chemical-enrichment history of the Galactic halo.
It will also support stellar-parameter calibration, studies of rare objects and key evolutionary phases, and time-domain spectroscopy of binary systems.

These goals require broad sky coverage, high-quality spectra for bright sources, and dedicated target lists for populations that cannot be selected efficiently using broad photometric criteria alone.
Accordingly, LAMOST~III LRS consists of three components: a main survey, 35 add-on catalogs, and a time-domain survey.
The following subsections describe the main scientific motivations and observational requirements of these components.

\subsection{The Galactic Disk}

The formation and evolution of the Galactic disk remain central themes in Galactic archaeology.
Although the broad distinction between thin- and thick-disk populations is well established \cite{thick_disk1983,bensby2003,rix2013,bland2016}, the disk's global structure is still not fully mapped.
Key open questions include the extent and origin of outer-disk truncation \cite{van2001,kregel2004,van2007}; the spatial and age dependence of the warp and flaring \cite{momany2006,kalberla2008,uppal2024}; the disk's dynamical response to the Galactic bar \cite{dehnen2000,athanassoula2003,athanassoula2013,antoja2014}, spiral arms \cite{sellwood2002,minchev2010,rok2012,kawata2014,dobbs2014}, and satellite perturbations (e.g., the Sagittarius dwarf; \cite{widrow2012,purcell2011,laporte2019L}).
Addressing these issues is particularly challenging beyond a Galactocentric radius of $\sim 15\,\mathrm{kpc}$, where spectroscopic coverage remains sparse.

LAMOST~III LRS will substantially enlarge the spectroscopic sample of distant giants, subgiants, main-sequence turn-off stars, and early-type stars.
Together with \emph{Gaia} astrometry and distances and complementary photometric surveys, the resulting radial velocities and stellar parameters will enable comparisons of the spatial, chemical, and kinematic distributions of disk populations across evolutionary stages and age ranges.
The improved sampling of bright giants and other distant tracers will sharpen maps of the outer disk, test whether flaring depends on stellar age or chemistry, and constrain the origin and dynamical evolution of the warp.

The disk also hosts spiral arms and star-forming regions, where young stellar populations interact with gas and dust.
Most current constraints on these regions come from infrared, millimeter, and radio surveys that primarily trace dust and molecular gas, leaving the stellar content, chemical properties, and detailed kinematics less well characterized.
Key questions include how turbulence, gas contraction, stellar feedback, and external perturbations regulate star formation \cite{kennicutt2012}.
Young stellar objects, OB and other massive stars, and diffuse interstellar bands provide complementary tracers of spiral-arm structure, recent star formation, and interstellar material.
LAMOST~III LRS will therefore build large spectroscopic samples of young and early-type stars, supplemented by dedicated add-on catalogs.
These data will enable studies of spiral-arm geometry; the spatial distribution and kinematics of young populations; and the link between recent star formation and the structure of the interstellar medium.

\subsection{The Galactic Halo and the Early Milky Way}

The Galactic halo, including both the stellar halo and the dark matter halo, is a fundamental component of the Milky Way. 
The stellar halo preserves evidence of the merger and accretion events that contributed to the assembly of the Galaxy, while the dark matter halo dominates the Galactic mass budget and shapes its dynamical evolution \cite{helmi2008,lucia2008,stewart2008}.
In recent years, Gaia and spectroscopic surveys such as APOGEE \cite{apogee2016}, H3 \cite{H32019}, the Southern Stellar Stream Spectroscopic Survey ($S^5$; \cite{S5}), and LAMOST I and II have revealed major structures such as Gaia-Sausage-Enceladus and have substantially improved our view of the inner halo \cite{Belokurov2018,helmi2008,mackereth2019,matteo2019}.
However, the outer halo remains less well characterized because distant tracers with spectroscopy, reliable distances, and chemical information are comparatively rare.

LAMOST~III LRS will increase both the number and spatial coverage of halo tracers, including red giants, subgiants, main-sequence turn-off stars, RR~Lyrae stars, and blue horizontal-branch stars.
Combined with \emph{Gaia} astrometry, these data will enable detailed studies of halo substructure and accretion debris, as well as the kinematic and metallicity distributions of halo populations.
The expanded distant-tracer sample will also provide key constraints for dynamical models of the Galactic potential and mass distribution.

Very metal-poor stars (VMP; $\mathrm{[Fe/H]}<-2$) provide a complementary window on the early Milky Way.
They preserve signatures of the earliest chemical-enrichment events, while extremely metal-poor stars (EMP; $\mathrm{[Fe/H]}<-3$) place unique constraints on the first generation of stars, their mass distribution, and the initial enrichment of the Galaxy \cite{beers2005,frebel2015,Klessen2023}.
LAMOST~III LRS will use candidate catalogs from photometric and spectrophotometric surveys to increase the efficiency of VMP/EMP observations.
It is expected to add $\sim 2\times 10^4$ VMP stars and $\sim 10^3$ EMP stars to existing spectroscopic samples, and to identify promising EMP candidates for high-resolution follow-up.
These observations will substantially improve constraints on early nucleosynthesis, the first stages of Milky Way formation, and the chemical evolution of the Galactic halo.

\subsection{Calibration Samples and Stellar Astrophysics}

Reliable stellar parameters and chemical abundances---and the masses and ages inferred from them---are essential for both disk and halo science.
LAMOST~I/II delivered very large stellar samples, but the corresponding calibration sets provided limited coverage of the Hertzsprung--Russell (HR) diagram, parameter space, and evolutionary phases.
LAMOST~III LRS will address this limitation by systematically observing benchmark and training samples, including \emph{Gaia} cluster members, wide binaries, eclipsing binaries, stars with high-resolution spectroscopic measurements, and stars with asteroseismic constraints.
It will also incorporate calibration sources from external databases and surveys such as PASTEL \cite{pastel2010,pastel2016} and SAGA \cite{saga2008,saga2011,saga2013,saga2017}, together with asteroseismic targets from \emph{Kepler} \cite{kepler2010} and TESS \cite{tess2015}.
These samples will improve calibration and validation of the LAMOST stellar-parameter pipelines across a broad range of HR-diagram locations, evolutionary states, and spectral types, and will support improved mass and age estimates for Galactic-archaeology applications.

The survey will also enhance the discovery and characterization of rare stellar populations and short-lived evolutionary phases.
The main survey increases sampling in sparsely populated regions of the Hertzsprung--Russell diagram, while dedicated add-on catalogs target classes that cannot be selected efficiently from HR-diagram position alone.
These samples include hot subdwarfs, Be stars, high-mass X-ray binaries, stars below the main sequence, cool dwarfs, high-proper-motion stars, variable stars, X-ray sources, and planet-host stars.
The resulting spectra will enable targeted studies of stellar evolution, compact systems, stellar activity, and sources of unusual SEDs.

\subsection{Time-Domain Spectroscopy}

Multi-epoch spectroscopy is essential for detecting radial-velocity variability, determining binary orbital parameters, and characterizing peculiar binary systems.
The LAMOST~III LRS time-domain survey will therefore obtain repeated observations of binaries and candidate peculiar systems.
It is designed to collect $\sim 2.15\times 10^5$ spectra for $\sim 1.2\times 10^4$ binaries, enabling binary-population studies, tests of binary-evolution pathways, and empirical constraints for population-synthesis models.
More details are described in Sec. \ref{sec:td}.

\section{Main Survey} \label{subsec:HRD}
LAMOST~III LRS is organized into three components: the main survey, a set of add-on catalogs, and a time-domain (TD) survey.
Each target is assigned an observing priority, where smaller values indicate higher priority. The main survey uses priorities $\geq 4$, whereas most add-on and TD targets are assigned priorities $\leq 3$ so that the highest-value objects are preferentially observed.
Targets are further split into two plate types by Gaia $G$ magnitude: the very-bright (VB) plate ($9 < G < 14$) and the bright-medium (BM) plate ($14 < G < 17.8$).

In this section, we describe the input catalog, target-selection strategy, and sampling performance of the main survey.

\subsection{Input Catalog}
The input catalog was constructed from Gaia DR3 \cite{2021A&A...649A...1G}. For stars with Gaia BP/RP slitless spectra ($G<17.65$), we derived their atmospheric parameters ($T_{\mathrm{eff}}$, $\log g$, [Fe/H], [$\alpha$/Fe], [C/Fe]) using the \texttt{DD-Payne} \cite{2019ApJS..245...34X}, trained on the LAMOST stellar parameter catalog of Xiang et al.~\cite{2019ApJS..245...34X}. The interstellar extinction $E(B-V)$ was simultaneously determined by fitting the Gaia BP/RP spectra. In addition, the absolute magnitude $M_G$ (and thus distance) was derived from the Gaia BP/RP spectra, astrometric parallaxes, and photometry, using a data-driven XGBoost regression algorithm \cite{Chen2016XGBoostAS}. A detailed description and validation of these stellar parameters derived from Gaia BP/RP spectra will be presented elsewhere (Xiang et al., in prep.).

Using the derived stellar parameters and extinction, we constructed a 3D dust map based on the best-measured stars, and inferred $M_G$ and $E(B-V)$
estimates for additional objects lacking Gaia BP/RP spectra, down to $G = 17.8$. Based on the $E(B-V)$ and the reddening coefficient $R_{BP-RP}$
from Zhang and Yuan~\cite{zhang2023}, we derived the intrinsic colours $(BP-RP)_0$
for individual Gaia sources. 
The reliability of the derived stellar parameters is further assessed through comparisons with independent measurements, as presented in \ref{sec:appendix}.
For this work, we discarded Gaia sources without parallax measurements or without $G$, $BP$, or $RP$ photometry from Gaia DR3, and applied a declination cut of $\mathrm{DEC} > -10^\circ$. Owing to its high completeness and precision, this catalog provides a solid basis for spectroscopic target selection. 

\subsection{Target Selection Strategy}

We partition the input catalog into approximately equal-area spatial bins of $\sim 1\,\mathrm{deg}^2$.
Sources are grouped in $1^\circ$ declination strips, with the right-ascension bin width scaled by $1/\cos(\mathrm{DEC})$ to compensate for projection effects.

For each bin, sources in special regions of the HR diagram are assigned the highest priority (4): those with $M_G > 12$, $M_G < -5$, or $(BP-RP)_0 < -0.5$.
We further split sources with $-5 < M_G < 12$ and $(BP-RP)_0 > -0.5$ into a moderately red subset ($-0.5 < (BP-RP)_0 < 3.2$) and a very red subset ($(BP-RP)_0 > 3.2$).
Within the very red subset, the brightest $50-a$ stars are assigned priority 4, where $a$ is the number of stars already assigned priority 4 by the criteria above.

For all remaining sources (i.e., those not assigned priority~4 by the criteria above),
we perform weighted random sampling independently within each spatial bin.

In each round, we select 200 targets, matching the $\sim 200$ fibers available per $\sim 1\,\mathrm{deg}^2$ LAMOST field.
The sampling weight for each source is defined as the product of three terms: an HRD weight, a $G$-magnitude weight, and an enhancement factor.

The HRD weight is taken as the inverse of the local density in the HR diagram, thereby upweighting rare or scientifically valuable populations (e.g., giants, M dwarfs, binaries, and turn-off stars).
To estimate the local density, we bin the HR diagram on a $37\times 85$ grid (0.2~mag in $M_G$ over $-5$ to 12, and 0.1~mag in $(BP-RP)_0$ over $-0.5$ to 3.2) and apply a 2D Gaussian smoothing kernel with $\sigma=2$ to reduce sensitivity to extremely dense regions.

The $G$-magnitude weight is defined as the inverse of the local density in the $G$ distribution, evaluated in 0.2~mag bins.
This term balances the sampling of bright and faint stars and indirectly boosts the overall signal-to-noise ratio, because brighter targets yield higher-SNR spectra.

\begin{figure}[H]
    \centering
    \includegraphics[width=1\linewidth]{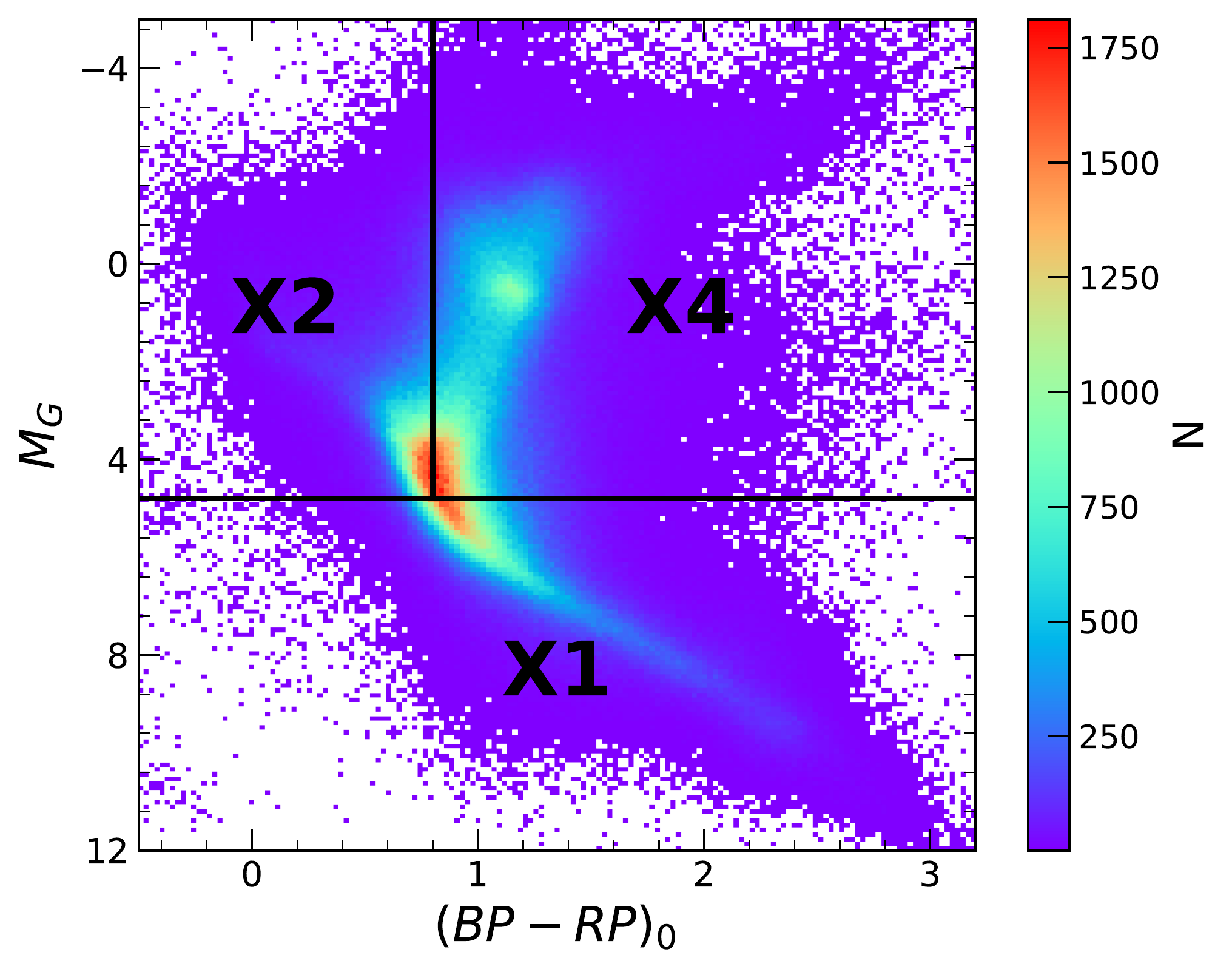}
    \caption{Enhancement factors adopted in the main-survey target-selection strategy. The background shows the number-density distribution of reddening-corrected sources from the input catalog within $28^\circ \leq \mathrm{Decl.}\leq29^\circ$ in the H-R diagram. The black lines at $M_G=4.8$ and $(BP-RP)_0=0.8$ divide the diagram into three regions, with enhancement factors of 4 for $M_G\leq 4.8$ and $(BP-RP)_0>0.8$, 2 for $M_G\leq 4.8$ and $(BP-RP)_0\leq 0.8$, and 1 otherwise.}
    \label{fig:enhancement_factor}
\end{figure}

We apply an enhancement factor to further upweight giants, turn-off stars, and young stars.
Specifically, sources with $M_G\leq 4.8$ receive a factor of 4 if $(BP-RP)_0>0.8$ and a factor of 2 if $(BP-RP)_0\leq 0.8$; all other sources receive a factor of 1.
These regions and enhancement factors are illustrated in Figure~\ref{fig:enhancement_factor}.

Finally, we carry out $N$ successive rounds of weighted random sampling without replacement within each spatial bin, separately for the VB and BM plates, using the combined weight defined above.
The number of sampling rounds, $N$, is determined by the source number in each spatial bin, ensuring that every star is ultimately assigned an observing priority.
Targets selected in the first round are assigned priority 4; those selected in the second round are assigned priority 5; and the priority increases by 1 in each subsequent round.
To maximize the inclusion of giants and turn-off stars, if an round selects fewer than 50 sources with $M_G\leq 4.8$ and $(BP-RP)_0\geq 0.8$, we instead include all such sources in that bin.
If a VB plate contains fewer targets than available fibers, the remaining fibers are used for supplementary observations of stars with $14 < G < 15$.
To favor new targets, sources that have already been observed by LAMOST DR10 with $\mathrm{S/N}_g>20$ have their priority value increased by 10.
Apart from this adjustment for previous LAMOST observations, we do not account for whether targets have been observed by other spectroscopic surveys.
Consequently, in areas of overlapping sky coverage, some selected targets may also have spectra available from external surveys.
These overlapping sources enable cross-survey consistency checks and, when observed at multiple epochs, provide a longer temporal baseline for studies of spectroscopic variability.

\subsection{Results for the Main Survey}
\begin{figure}[H]
    \centering
    \includegraphics[width=1\linewidth]{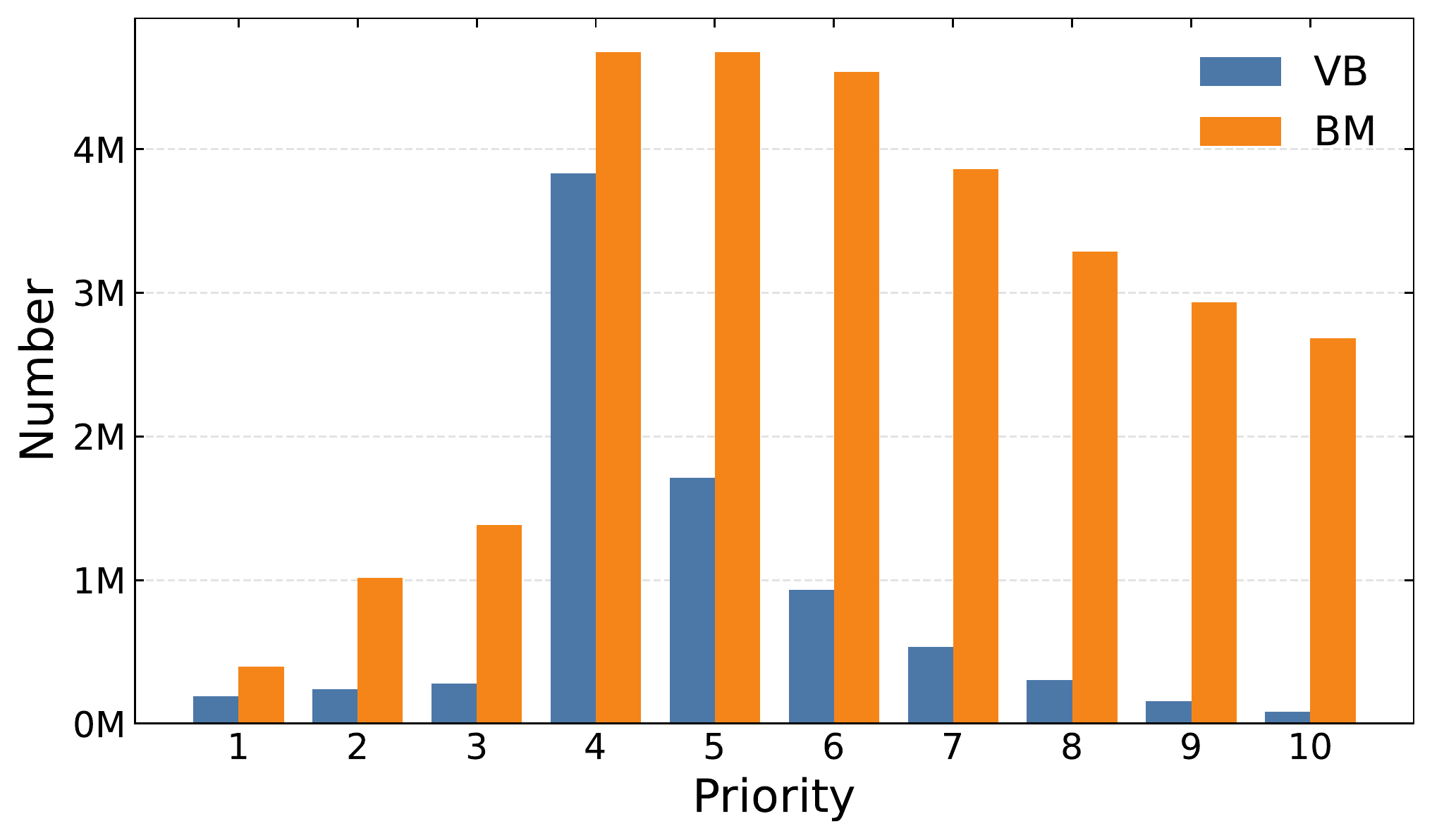}
    \caption{Distribution of the top ten priority classes for sources of LAMOST~III LRS in VB and BM plates.}
\label{fig:main_survey_pri}
\end{figure}

\begin{figure*}
    \centering
    \includegraphics[width=\textwidth]{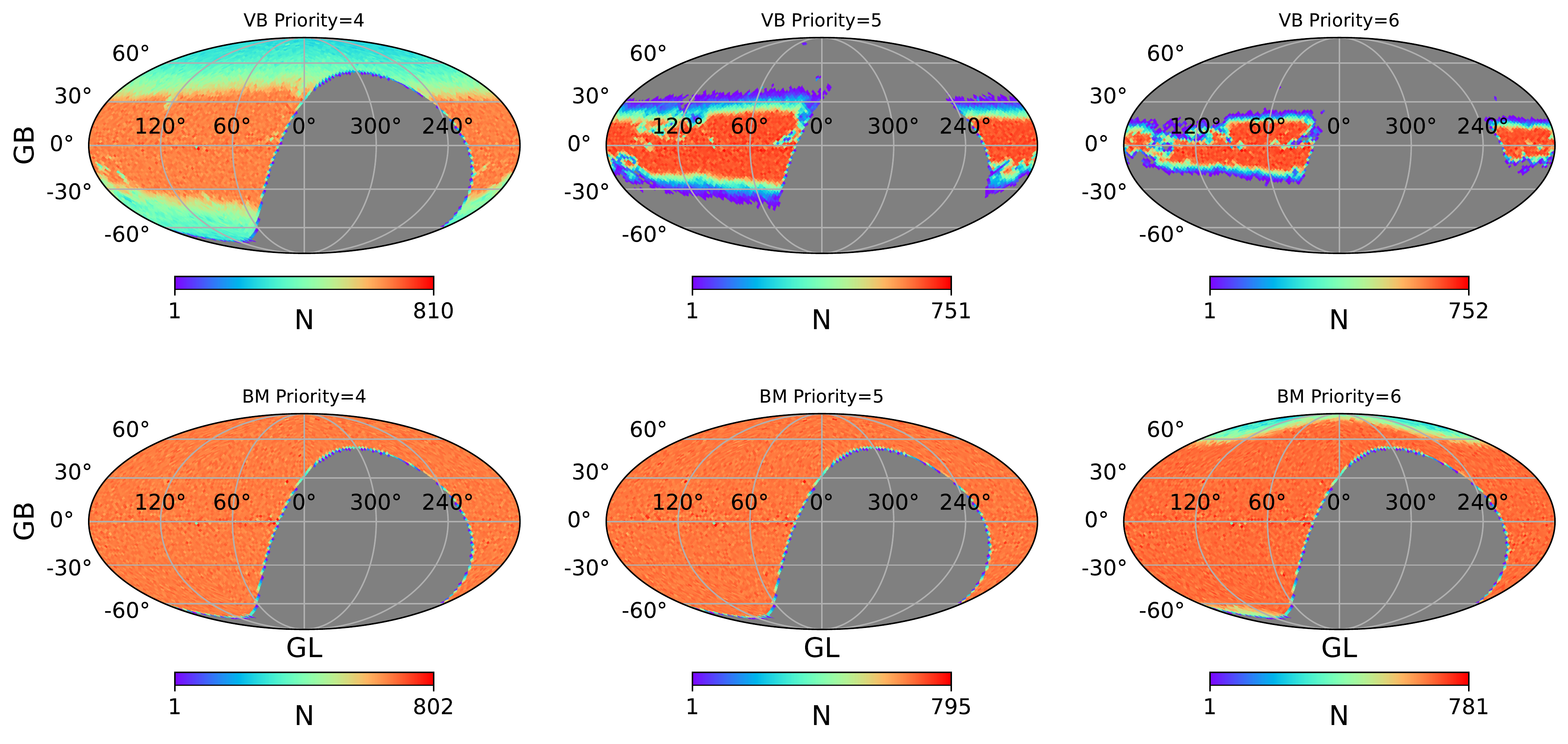}
    \caption{Spatial distribution of main-survey sources with priorities 4--6 in the VB (top) and BM (bottom) plates, color-coded by number density.}
    \label{fig:mian_survey_space}
\end{figure*}

The main survey assigns priorities to approximately 100 million sources.
The priority distribution for all LAMOST~III LRS targets is shown in Figure~\ref{fig:main_survey_pri}.
For the VB plates, the number of priority-5 (and lower-priority) targets is significantly smaller than for the BM plates.
This is because, at high Galactic latitudes, the source density for the VB plates is too low to accommodate two or more complete sampling rounds, as illustrated in the top panel of Figure \ref{fig:mian_survey_space}.

\begin{figure*}
    \centering
    \includegraphics[width=\textwidth]{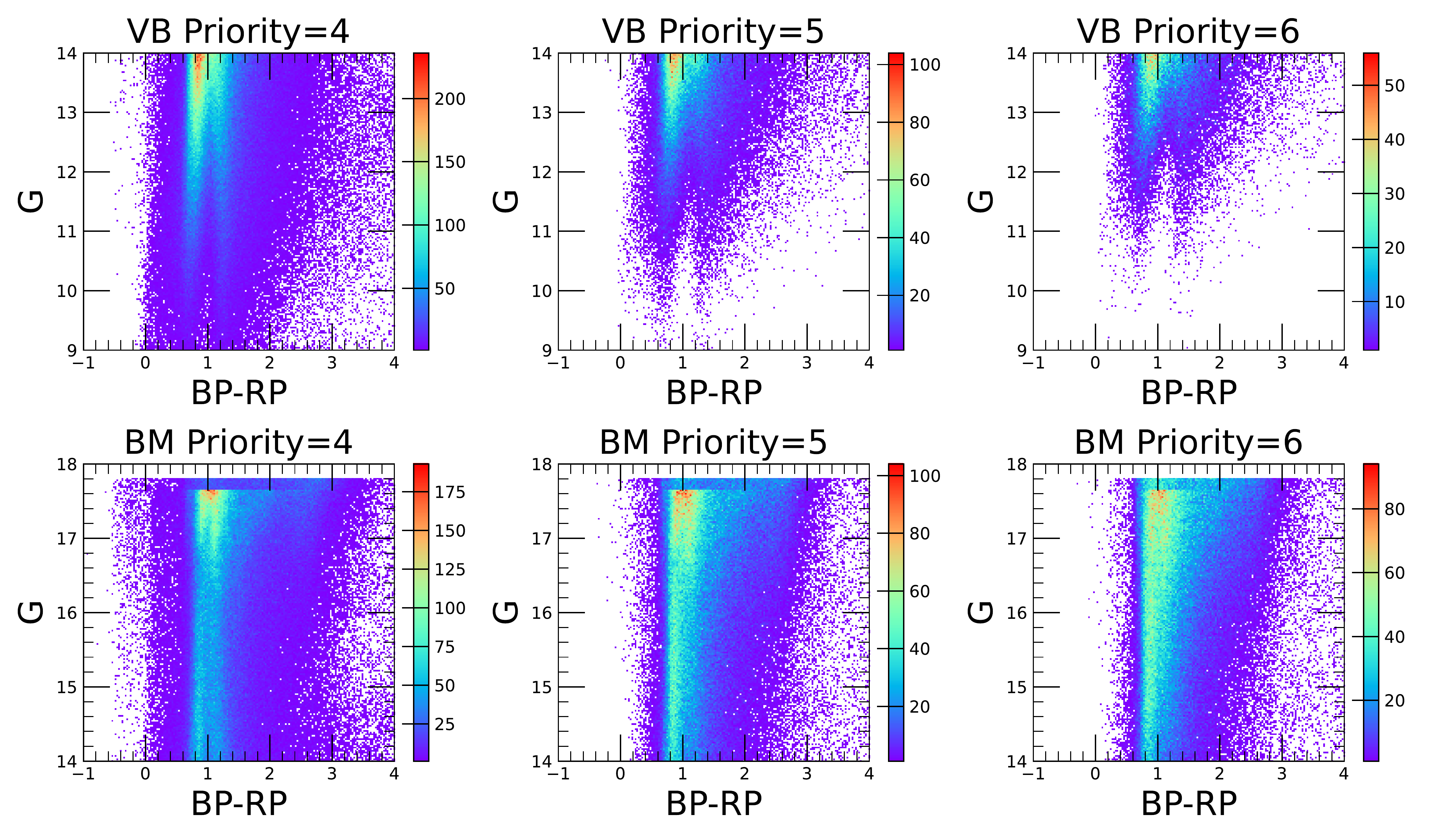}
    \caption{Color--magnitude distribution of main-survey sources with priorities 4--6 in the VB (top) and BM (bottom) plates, color-coded by number density.}
    \label{fig:mian_survey_cmd}
\end{figure*}

The spatial and color--magnitude distributions of the three highest-priority main-survey samples are shown in Figures \ref{fig:mian_survey_space} and \ref{fig:mian_survey_cmd}, respectively.
As shown in Figure \ref{fig:mian_survey_space}, VB-plate sources with priority $\geq 5$ occur only at low and intermediate Galactic latitudes.
By contrast, BM-plate sources with priorities 4--6 are distributed nearly uniformly across Galactic coordinates, with a modest enhancement toward the Galactic disk.

In the color--magnitude diagram (Figure \ref{fig:mian_survey_cmd}), dwarfs and giants form two clearly separated branches, with a more pronounced split for the VB plates.
For the VB plates, the number of bright priority-6 sources is smaller than that of priority-4 sources, reflecting the combined effects of larger $G$-band weights at bright magnitudes and the low source density at high Galactic latitudes, which prevents two complete sampling rounds.

\begin{figure*}
    \centering
    \includegraphics[width=\textwidth]{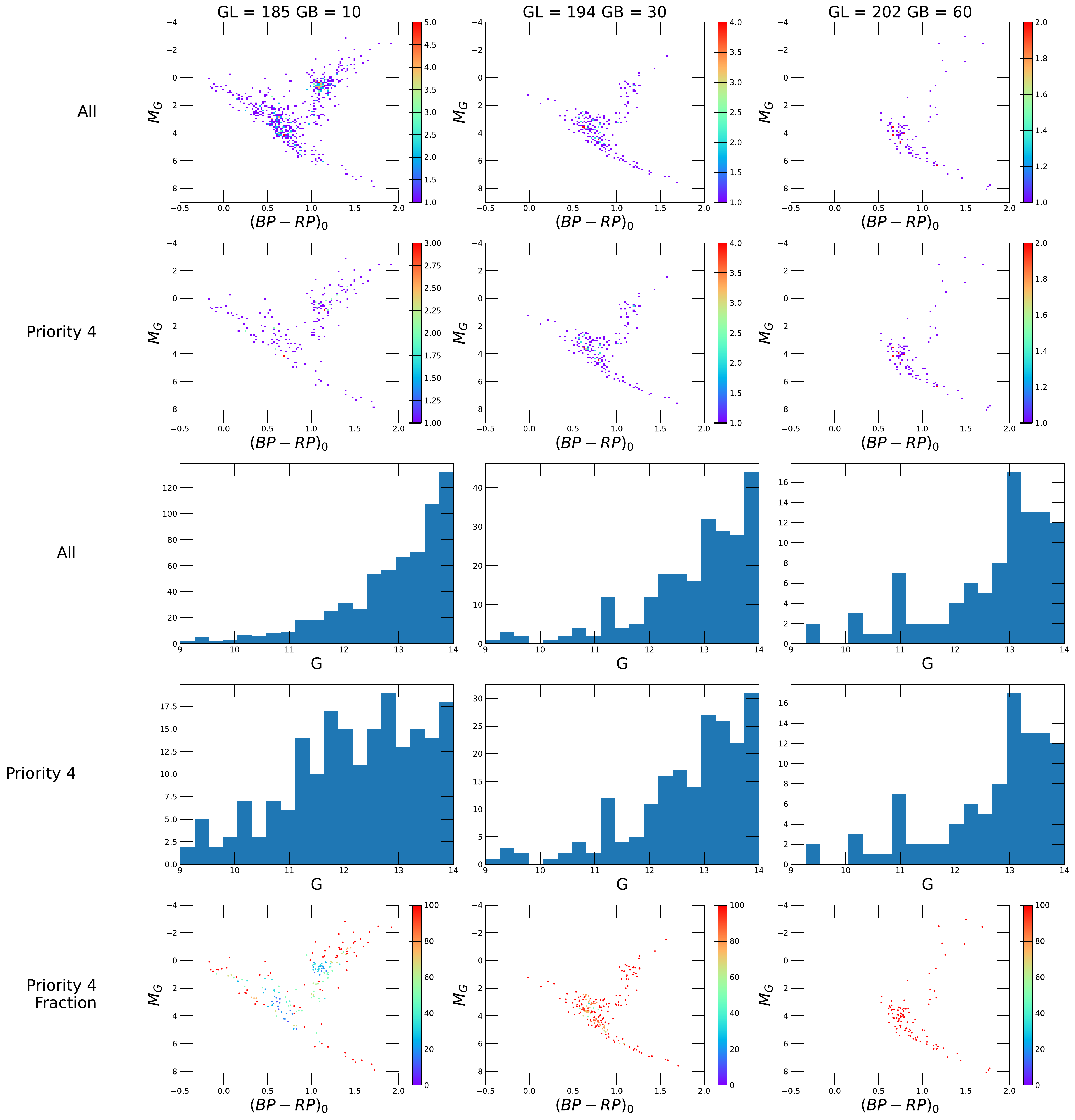}
    \caption{Comparison between the input catalog and the priority-4 sample in three spatial bins at Galactic latitudes ($b=10^\circ$ left, $30^\circ$ middle, and $60^\circ$ right) for the main survey in the VB plates. Top two rows: input-catalog and priority-4 distributions in the HR diagram, color-coded by number density. Middle two rows: input-catalog and priority-4 distributions in $G$ magnitude.
    Bottom row: sampling fraction of the priority-4 sample in the HR diagram.}
    \label{fig:mian_survey_point_vb}
\end{figure*}

\begin{figure*}
    \centering
    \includegraphics[width=16cm]{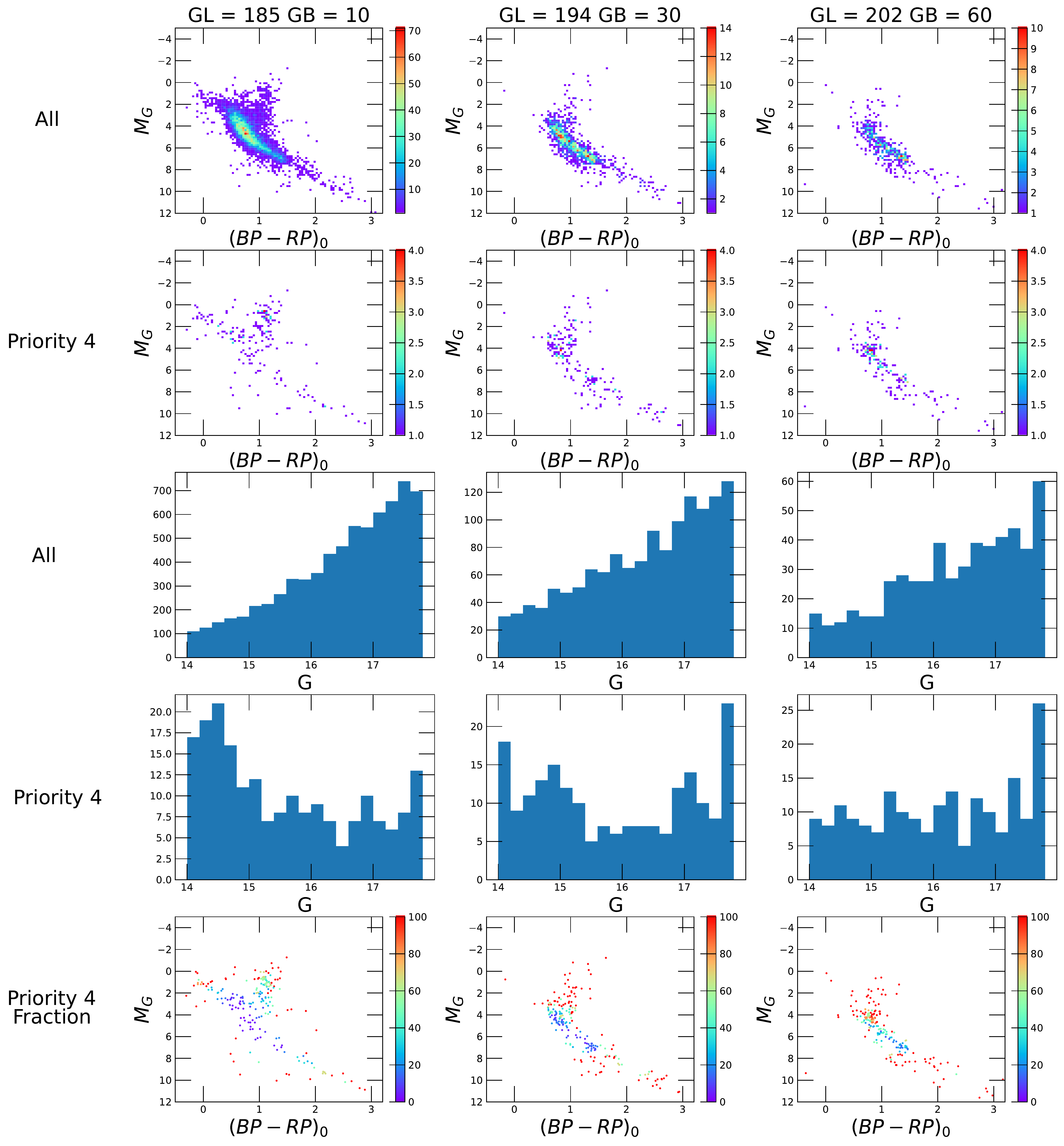}
    \caption{Same as Figure \ref{fig:mian_survey_point_vb}, but for the BM plates.}
    \label{fig:mian_survey_point_bm}
\end{figure*}

We select three spatial bins at Galactic latitudes $b=10^\circ$, $30^\circ$, and $60^\circ$ to illustrate the main-survey results for the VB and BM plates (Figures \ref{fig:mian_survey_point_vb} and \ref{fig:mian_survey_point_bm}).
The top two rows show the effect of the HRD weights: compared to the raw sample, priority-4 sources are distributed more uniformly across the HRD.
The middle two rows show the effect of the $G$-band weights, which yield an approximately uniform $G$-magnitude distribution after the first sampling.
This effect is less apparent for the VB plates because the first sampling already includes essentially all sources at intermediate and high Galactic latitudes.
The bottom rows show the first-sampling ratio across the HRD, demonstrating that low-density regions are sampled at substantially higher rates.
At high Galactic latitude, nearly all VB-plate sources are assigned priority 4 because the source density is too low to support two complete sampling rounds.

This target-selection strategy adopts a non-uniform target-selection strategy that preferentially selects specific stellar populations (e.g., giants and turn-off stars). 
Selection effects should therefore be considered when using the spectroscopic sample for statistical studies of Galactic stellar populations and structure. 
They can be quantified by comparing the color--magnitude distributions of the spectroscopic sample and the underlying Gaia~DR3 photometric parent sample, while accounting for spatial variations and the completeness of stellar-parameter measurements; see \cite{liu2017,Chen2018}.

\section{Add-on Catalogs and TD survey}\label{sec:add_on}

In addition to the main survey, LAMOST~III LRS includes 35 add-on catalogs and the TD survey.
Duplicate sources may occur between the main survey, the add-on catalogs, and the TD survey (and also among the add-on catalogs); in such cases, we retain the highest assigned priority.

\subsection{Add-on Catalogs}

The add-on catalogs comprise high-priority samples that complement the main survey and improve the sampling of scientifically valuable, relatively rare objects.
Most add-on targets are assigned higher priorities than main-survey targets, except for a small fraction of sources in the Kepler and Local Cool Dwarf samples.
In general, smaller catalogs are assigned higher priorities to increase observing completeness.
We also simulate realistic observations to estimate the expected observing fraction of each add-on catalog.
The simulations use the LAMOST survey-strategy system, which optimizes sky tiling subject to telescope-pointing constraints and allocates fibers to targets based on their priorities and fiber-positioning constraints \cite{yuan2012,yuan2014}.

\begin{figure*}
    \centering
    \includegraphics[width=\textwidth]{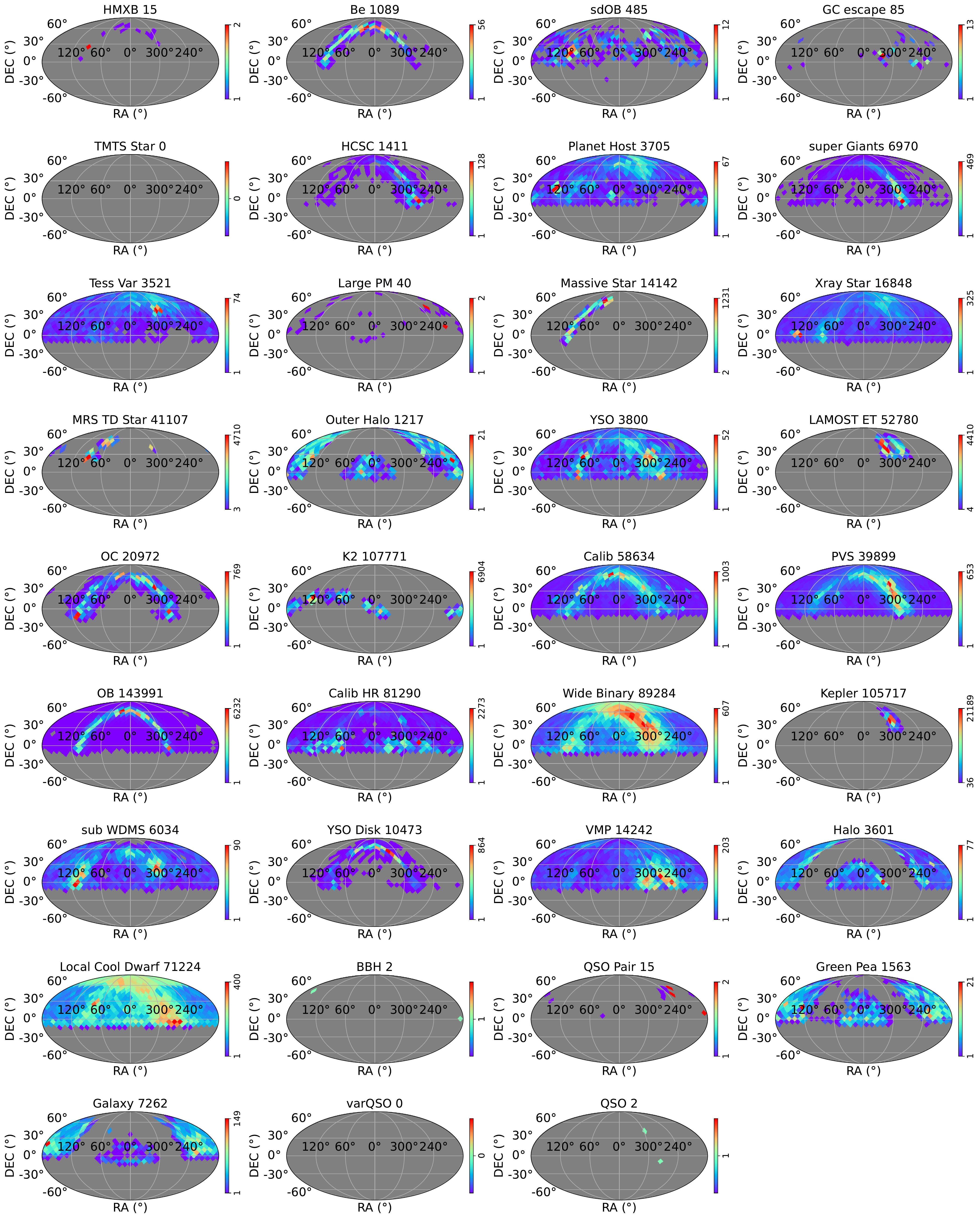}
    \caption{Spatial distribution of add-on targets in the VB plates, color-coded by number density. The panel titles list the sample name and the number of targets in the VB plates.}
    \label{fig:add_on_vb}
\end{figure*}

\begin{figure*}
    \centering
    \includegraphics[width=\textwidth]{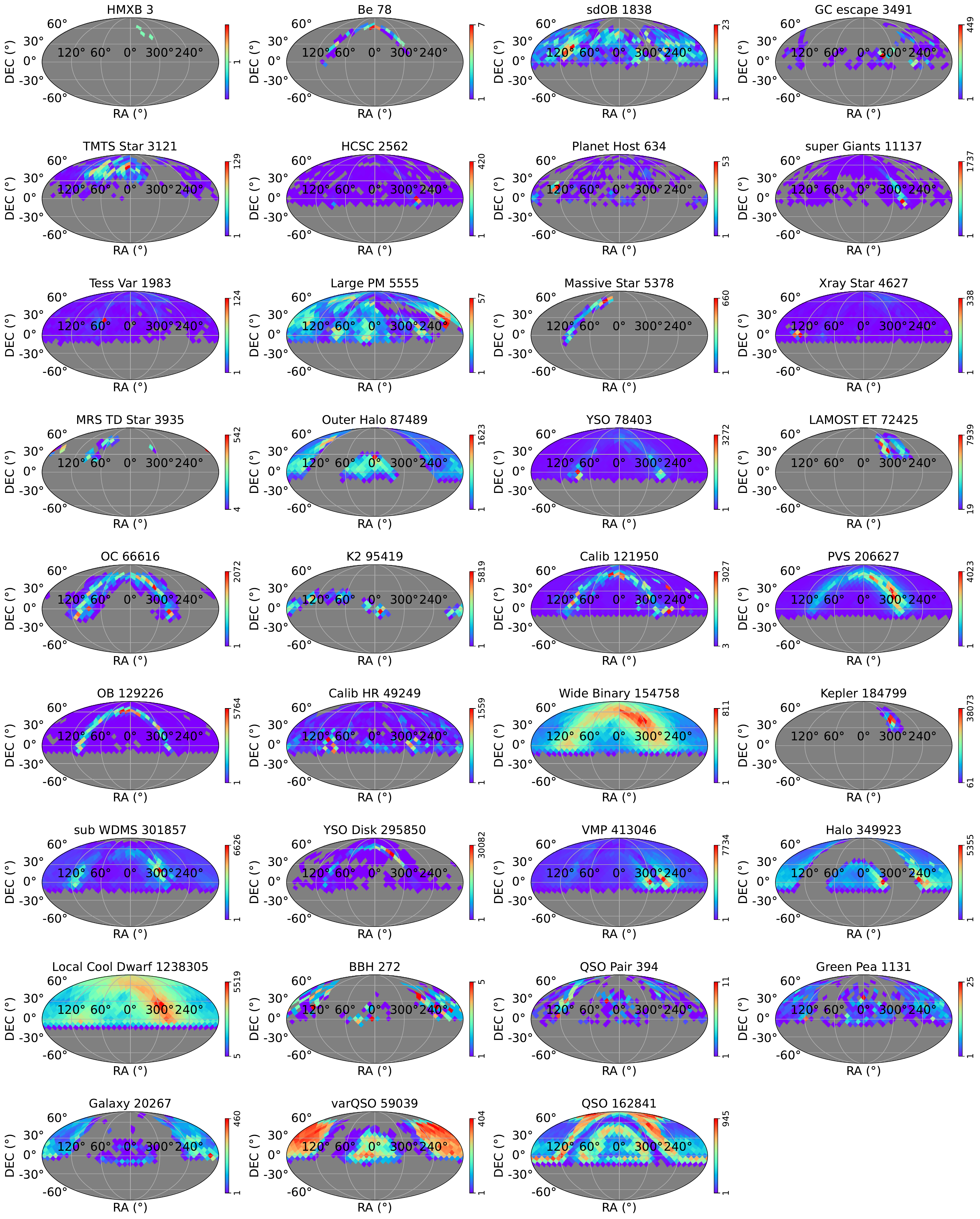}
    \caption{Same as Figure \ref{fig:add_on_vb}, but for the BM plates.}
    \label{fig:add_on_bm}
\end{figure*}

The add-on catalogs are divided into three groups: Milky Way~I (10 catalogs), Milky Way~II (19 catalogs), and extragalactic (6 catalogs).
The spatial distribution and the number of add-on targets on the VB and BM plates are shown in Figures \ref{fig:add_on_vb} and \ref{fig:add_on_bm}, respectively.
Table \ref{table1} summarizes each catalog's description, priority, number of sources, and simulated observing fraction.
As shown in Table \ref{table1}, smaller catalogs are typically observed at higher fractions.
However, the observing fraction can be substantially reduced for catalogs that are strongly clustered on the sky, because only a limited number of fibers can be allocated within a given field.

The Milky Way~I samples comprise stars selected by their location in the HR diagram.
They include halo and outer-halo stars, supergiants, O- and B-type stars, massive stars, young stellar objects (YSOs), stars below the main sequence, and M-type dwarfs.
Because these objects also receive higher priority in the main survey based on their HR-diagram positions, their simulated observing fractions are generally higher than those of larger, more broadly distributed catalogs, even though several Milky Way~I samples contain more than $10^5$ sources.
Most Milky Way~I samples concentrate on the Galactic disk, except for the halo and outer-halo catalogs, as shown in Figures \ref{fig:add_on_vb} and \ref{fig:add_on_bm}.

The Milky Way~II samples comprise targets selected largely independent of their positions in the HR diagram.
They include very metal-poor (VMP) stars, peculiar variables, binaries, calibration stars, clusters, kinematically selected targets (e.g., high-proper-motion stars and stars escaping from globular clusters), sources from time-domain collaborations (e.g., ET, TMTS, Kepler, and the LAMOST MRS TD survey), exoplanet host stars, and X-ray sources.

The extragalactic samples include galaxies, quasi-stellar objects (QSOs), changing-look active galactic nuclei, Green Pea galaxies, QSO pairs, and candidates for supermassive binary black holes.
Most extragalactic targets are assigned to the BM plates because galaxies are typically fainter; Green Pea galaxies are a notable exception.

\setlength{\tabcolsep}{1.2mm}{
\begin{table*}[htbp]
\scriptsize
\centering
\caption{Descriptions and simulated observing fractions for the add-on catalogs. Entries in the priority columns give the number of sources in the VB/BM plates, with the simulated observing fraction in parentheses.}
\resizebox{\textwidth}{!}{
\begin{tabular}{llllll}
\hline
\hline
Name & Pri 1 & Pri 2 & Pri 3 & All & Description \\
\hline
\hline
HMXB$^a$ & 15/3 (72.2\%) & - & - & 72.2\% & High-Mass X-ray Binary\\
Be$^a$ & 1089/78 (66.6\%) & - & - & 66.6\% & Rapidly rotating B-type stars\\
sdOB$^a$ & 485/1838 (63.0\%) & - & - & 63.0\% & O- and B-type hot subdwarfs\\
GC escape$^a$ & 85/3491 (51.0\%) & - & - & 51.0\% & Stars escaping from globular clusters\\
TMTS Star$^a$ & -/3121 (65.4\%) & - & - & 65.4\% & Contact binaries, stellar flares, and\\
 & & & & & cataclysmic variables from TMTS\\
HCSC$^a$ & 1411/2562 (62.7\%) & - & - & 62.7\% & Hyper-compact star clusters\\
Planet Host$^a$ & 3705/634 (63.2\%) & - & - & 63.2\% & Exoplanet host stars\\
Supergiants$^a$ & - & 6970/11,137 (63.9\%) & - & 63.9\% & Supergiants\\
TESS Var$^a$ & - & 2276/1424 (33.7\%) & 1245/559 (27.1\%) & 31.5\% & Variables from TESS\\
Large PM$^a$ & 40/5555 (55.2\%) & - & - & 55.2\% & High-proper-motion stars\\
Massive Star$^a$ & - & 9722/4577 (51.6\%) & 4420/801 (40.8\%) & 48.7\% & Massive stars\\
X-ray Star$^a$ & - & 11,621/3714 (56.0\%) & 5227/913 (59.2\%) & 57.0\% & X-ray sources\\
MRS TD Star$^a$ & 41,107/3935 (56.2\%) & - & - & 56.2\% & Targets in the MRS time-domain survey\\
Outer Halo$^a$ & 1217/87,489 (57.2\%) & - & - & 57.2\% & Outer-halo stars\\
YSO$^a$ & 278/7458 (58.0\%) & 3522/70,945 (51.5\%) & - & 52.1\% & Young stellar objects\\
LAMOST ET$^a$ & 43,897/21,465 (53.4\%) & 6404/20,587 (29.9\%) & 2479/30,373 (32.1\%) & 42.7\% & Targets in the ET survey\\
OC$^a$ & 3038/9017 (45.2\%) & 17,934/57,599 (42.7\%) & - & 43.0\% & Open clusters\\
K2$^a$ & 60,493/60,248 (53.4\%) & 47,278/35,171 (48.2\%) & - & 51.3\% & Targets in the K2 fields\\
Calib$^a$ & 12,945/20,784 (38.3\%) & 45,288/98,053 (32.8\%) & 401/3113 (8.3\%) & 33.4\% & Calibration stars from wide binaries\\
 & & & & & and clusters\\
PVS$^a$ & 314/4546 (67.1\%) & 22,513/52,285 (55.1\%) & 17,072/149,796 (42.3\%) & 46.7\% & Pulsating variable stars\\
OB$^a$ & 2014/2771 (67.1\%) & 52,996/31,410 (54.0\%) & 88,989/95,045 (37.1\%) & 42.9\% & O- and B-type stars\\
Calib HR$^a$ & 2632/1127 (60.6\%) & 3154/15,474 (18.6\%) & 75,504/32,648 (47.7\%) & 43.9\% & Calibration stars from high-resolution\\
 & & & & & spectroscopy\\
Wide Binary$^a$ & 6682/33,106 (67.4\%) & 8672/6165 (56.4\%) & 73,930/115,487 (46.5\%) & 50.5\% & Wide binaries\\
Kepler$^a$ & 5345/12,982 (68.3\%) & 18,273/86,948 (20.0\%) & 28,204/9503 (45.0\%) & 21.2\% & Targets in the Kepler field\\
 & \multicolumn{2}{l}{Pri 4: 1978/31,477 (8.9\%)} & \multicolumn{2}{l}{Pri 5: 51,974/43,889 (8.3\%)} & \\
sub WDMS$^a$ & 1559/22,028 (62.5\%) & 4475/65,269 (58.7\%) & -/214,560 (43.6\%) & 48.5\% & Sub-main-sequence stars\\
YSO Disk$^a$ & 796/5660 (60.5\%) & 3265/27,007 (49.1\%) & 6412/263,183 (16.4\%) & 20.6\% & Young stellar objects in the Galactic\\
 & & & & & disk\\
VMP$^a$ & 2121/31,649 (51.1\%) & 3344/69,784 (45.9\%) & 8777/311,613 (36.6\%) & 39.4\% & Very metal-poor (VMP) stars\\
Halo$^a$ & - & 3601/349,923 (53.0\%) & - & 53.0\% & Halo stars\\
Local Cool Dwarf$^a$ & 1/50 (74.5\%) & - & 43,966/221,765 (45.9\%) & 22.9\% & Local cool dwarfs\\
 & \multicolumn{2}{l}{Pri 5: 27,257/1,016,490 (17.0\%)} & & & \\
BBH$^b$ & 2/272 (28.8\%) & - & - & 28.8\% & Candidates for supermassive binary black holes\\
QSO Pair$^b$ & 15/394 (28.8\%) & - & - & 28.8\% & QSO pairs\\
Green Pea$^b$ & 1563/1131 (62.5\%) & - & - & 62.5\% & Green Pea galaxies\\
Galaxy$^b$ & 1797/5051 (59.7\%) & 5465/15,216 (50.1\%) & - & 52.5\% & Galaxies\\
varQSO$^b$ & -/904 (62.9\%) & -/58,135 (58.6\%) & - & 58.7\% & Changing-look AGN\\
QSO$^b$ & 2/60,059 (57.0\%) & -/61,309 (51.3\%) & -/41,473 (43.8\%) & 51.5\% & Quasi-stellar objects (QSOs)\\

\hline
\hline
\end{tabular}
}
\begin{tablenotes}
\footnotesize
\item $^a$ Milky Way samples.
\item $^b$ Extragalactic samples.
\end{tablenotes}
\label{table1}
\end{table*}}

\subsection{Time-domain (TD) Survey}\label{sec:td}

The TD survey is a high-priority program focused on binaries and other peculiar binary systems.
In its first year, it plans to observe three eROSITA Final Equatorial Depth Survey (eFEDS)\cite{brunner2022} plates, with eight visits per plate; each plate is expected to contain approximately 330 Galactic and 310 extragalactic X-ray sources, with $G=13.5$--17.5.
Over the five-year LAMOST~III LRS, the TD program will also observe ten additional low-Galactic-latitude plates, with twenty visits per plate.
These plates target O- and B-type stars, sub-main-sequence stars, and special subdwarf B (sdB) stars.
On average, each plate contains about 100 OB stars, 900 sub-main-sequence stars, and 5 sdB stars, with $G=14$--17.5.

These extensive time-domain spectroscopic observations will yield a large sample with well-determined parameters, providing key inputs for binary population synthesis and enabling stringent tests of binary-formation and evolution models.
In particular, the TD survey targets many peculiar binary systems, which can provide direct evidence for key evolutionary channels and place strong constraints on the underlying physical processes.

\section{Survey Status}\label{sec:now}

\begin{figure*}
    \centering
    \includegraphics[width=\textwidth]{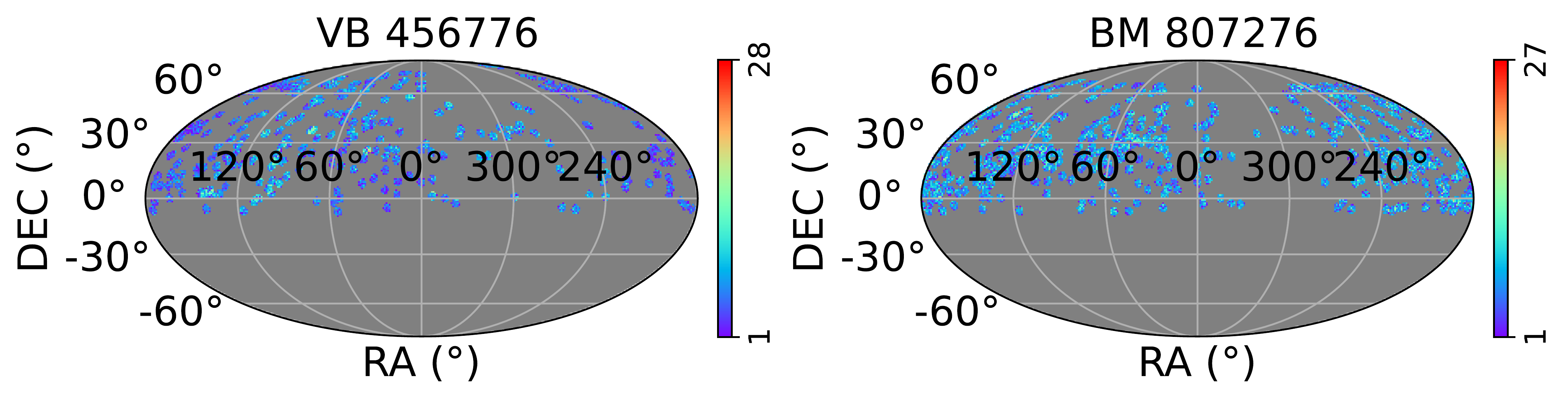}
    \caption{Spatial distribution of LAMOST~III LRS targets in the VB (left) and BM (right) plates, color-coded by number density. The panel titles report the number of targets.}
    \label{fig:space}
\end{figure*}

\begin{figure*}
    \centering
    \includegraphics[width=\textwidth]{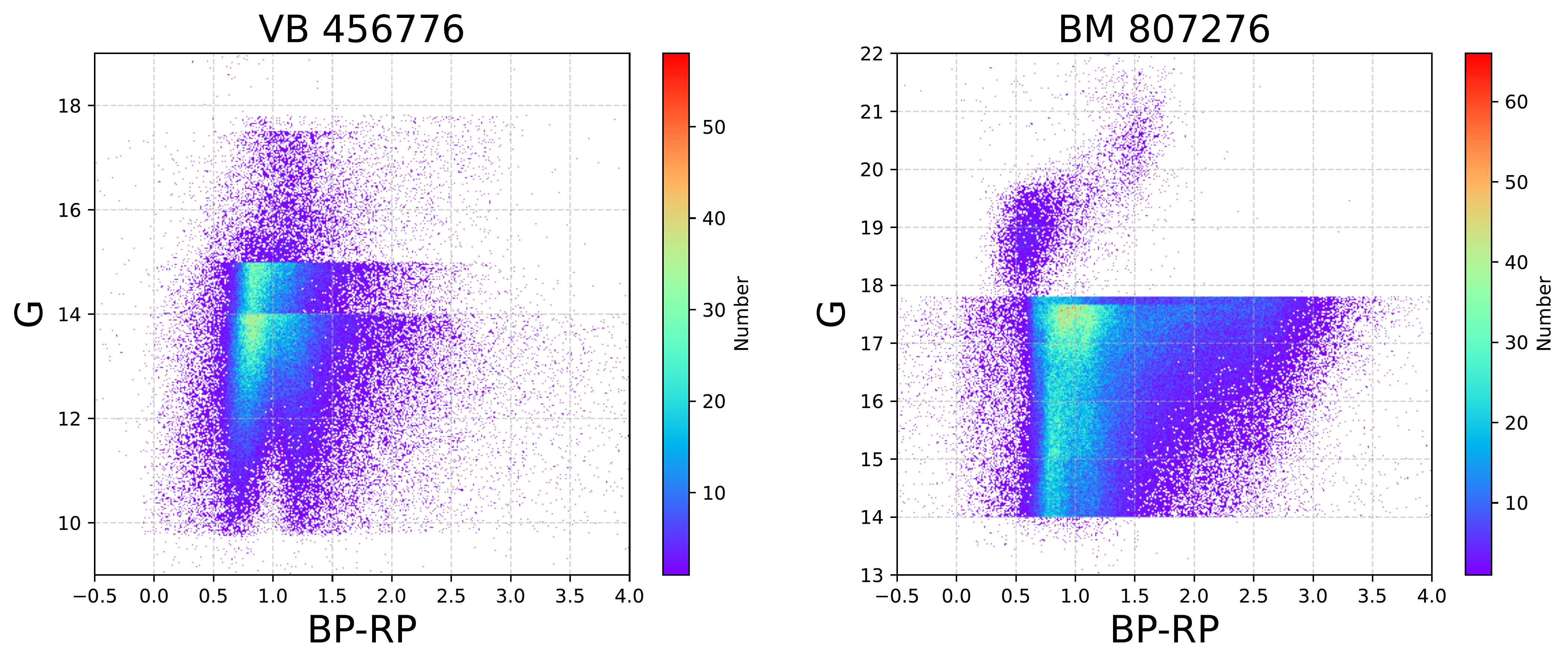}
    \caption{Color--magnitude distribution of LAMOST~III LRS targets in the VB (left) and BM (right) plates, color-coded by number density. The panel titles report the number of targets.}
    \label{fig:cmd}
\end{figure*}

To June 2025, LAMOST~III LRS has been operational for about two years and has released DR12 and DR13, covering 2023-09-18 to 2025-06-30. These releases provide 1.53 million low-resolution spectra from 658 plates and 4.19 million medium-resolution spectra from 1,011 plates.
In the following, we overview only the low-resolution spectroscopic survey.
After removing duplicates, the low-resolution sample contains 1,264,052 unique sources; 37\% of VB-plate observations are removed as duplicates.
Figures \ref{fig:space} and \ref{fig:cmd} show their spatial distribution and color--magnitude diagram.
As noted above, some sources with $14<G<15$ were observed on the VB plates when fibers were idle.
In Figure \ref{fig:cmd}, sources fainter than $G=18$ are primarily QSOs and other extragalactic targets.
For the VB/BM plates, 83.11\%/38.52\% of sources have $\mathrm{S/N}_g>20$, and 50.03\%/15.1\% have $\mathrm{S/N}_g>50$.
The fraction of high-quality data may be slightly underestimated because about 2,500 extragalactic galaxies observed on the BM plates are largely insensitive to the $g$-band S/N.
Main-survey targets account for 74.64\%/61.38\% of the observed sources in the VB/BM plates, respectively; the numbers for each add-on catalog are listed in Table \ref{table3}.

\setlength{\tabcolsep}{2.8mm}{
\begin{table*}[htbp]
\footnotesize
\caption{Number of observed sources in the add-on catalogs from 2023-09-18 to 2025-06-30.}
\begin{tabular}{llll}
\hline
\hline
\\
 &   \multicolumn{3}{c}{Number(Observed ratio) in VB/BM plate} \\
 \\
\cline{2-4}  \\
Name  & Pri 1 & Pri 2 & Pri 3 \\
\hline
\hline
HMXB$^a$ & 1(6.7\%)/- & - & - \\
Be$^a$ & 127(11.7\%)/14(17.9\%) & - & - \\
sdOB$^a$ & 73(15.1\%)/368(20.0\%) & - & - \\
GC escape$^a$ & 7(8.2\%)/220(6.3\%) & - & - \\
TMTS Star$^a$ & 36(\textemdash)/393(12.6\%) & - & - \\
HCSC$^a$ & 64(4.5\%)/167(6.5\%) & - & - \\
Planet Host$^a$ & 344(9.3\%)/147(23.2\%) & - & - \\
Supergiants$^a$ & - & 210(3.0\%)/104(0.9\%) & - \\
TESS Var$^a$ & - & 155(6.8\%)/126(8.8\%) & 102(8.2\%)/101(18.1\%) \\
Large PM$^a$ & 15(37.5\%)/967(17.4\%) & - & - \\
Massive Star$^a$ & - & 760(7.8\%)/978(21.4\%) & 404(9.1\%)/113(14.1\%) \\
X-ray Star$^a$ & - & 990(8.5\%)/402(10.8\%) & 570(10.9\%)/136(14.9\%) \\
MRS TD Star$^a$ & 5,531(13.5\%)/553(14.1\%) & - & - \\
Outer Halo$^a$ & 488(40.1\%)/16,107(18.4\%) & - & - \\
YSO$^a$ & 25(9.0\%)/717(9.6\%) & 341(9.7\%)/6,128(8.6\%) & - \\
LAMOST ET$^a$ & 3,369(7.7\%)/1,653(7.7\%) & 410(6.4\%)/656(3.2\%) & 162(6.5\%)/1,358(4.5\%) \\
OC$^a$ & 52(1.7\%)/1,104(12.2\%) & 840(4.7\%)/4,017(7.0\%) & - \\
K2$^a$ & 9,951(16.4\%)/16,286(27.0\%) & 8,812(18.6\%)/9,014(25.6\%) & - \\
Calib$^a$ & 930(7.2\%)/1,388(6.7\%) & 3,252(7.2\%)/5,262(5.4\%) & 39(9.7\%)/21(0.7\%) \\
PVS$^a$ & 31(9.9\%)/347(7.6\%) & 1,585(7.0\%)/3,921(7.5\%) & 2,625(15.4\%)/9,744(6.5\%) \\
OB$^a$ & 126(6.3\%)/150(5.4\%) & 4,719(8.9\%)/3,442(11.0\%) & 7,147(8.0\%)/5,917(6.2\%) \\
Calib HR$^a$ & 132(5.0\%)/119(10.6\%) & 110(3.5\%)/551(3.6\%) & 4,722(6.3\%)/3,667(11.2\%) \\
Wide Binary$^a$ & 769(11.5\%)/6,287(19.0\%) & 569(6.6\%)/561(9.1\%) & 5,473(7.4\%)/9,615(8.3\%) \\
Kepler$^a$ & 633(11.8\%)/781(6.0\%) & 1,156(6.3\%)/2,221(2.6\%) & 2,250(8.0\%)/230(2.4\%) \\
 & Pri 4: 63(3.2\%)/206(0.7\%) & Pri 5: 695(1.3\%)/105(0.2\%) & - \\
sub WDMS$^a$ & 404(25.9\%)/3,742(17.0\%) & 2,013(45.0\%)/6,895(10.6\%) & 2,598(\textemdash)/22,419(10.4\%) \\
YSO Disk$^a$ & 63(7.9\%)/395(7.0\%) & 179(5.5\%)/1,063(3.9\%) & 786(12.3\%)/7,205(2.7\%) \\
VMP$^a$ & 333(15.7\%)/4,660(14.7\%) & 295(8.8\%)/7,257(10.4\%) & 1,230(14.0\%)/29,910(9.6\%) \\
Halo$^a$ & - & 1,501(41.7\%)/55,130(15.8\%) & - \\
Local Cool Dwarf$^a$ & - & - & 3,827(8.7\%)/22,526(10.2\%) \\
 & Pri 5: 4,634(17.0\%)/32,363(3.2\%) & - & - \\
BBH$^b$ & -/54(19.9\%) & - & - \\
QSO Pair$^b$ & 1(6.7\%)/44(11.2\%) & - & - \\
Green Pea$^b$ & 2(0.1\%)/125(11.1\%) & - & - \\
Galaxy$^b$ & -/823(16.3\%) & -/1,871(12.3\%) & - \\
varQSO$^b$ & -/147(16.3\%) & 69(\textemdash)/9,002(15.5\%) & - \\
QSO$^b$ & 28(1400.0\%)/4,320(7.2\%) & 56(\textemdash)/4,235(6.9\%) & 30(\textemdash)/2,638(6.4\%) \\

\hline
\hline
\end{tabular}
\begin{tablenotes}
\footnotesize
\item $^a$ Milky Way samples.
\item $^b$ Extragalactic samples.
\item Note: when there are insufficient VB targets to fill a plate, brighter BM targets ($14<G<15$) are assigned to the remaining fibers. 
\item Thus, the VB counts in this table may include targets classified as BM in Table~\ref{table1}.
\end{tablenotes}
\label{table3}
\end{table*}}

\begin{figure*}
    \centering
    \includegraphics[width=\textwidth]{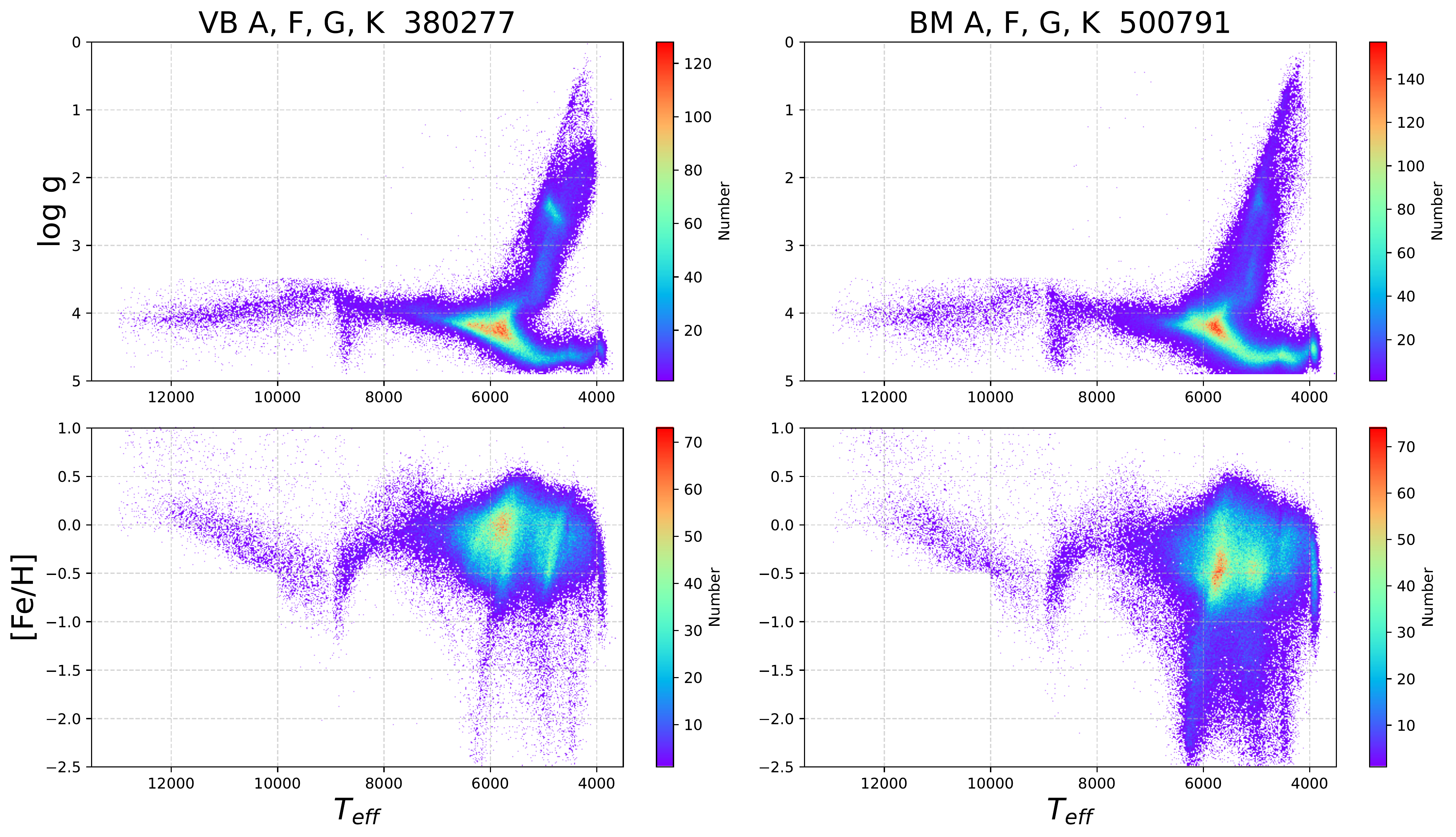}
    \caption{Distributions of A-, F-, G-, and K-type stars observed by LAMOST~III LRS in the HR diagram (top) and the $T_{\rm eff}$--[Fe/H] plane (bottom) for the VB (left) and BM (right) plates, color-coded by number density. The panel titles report the number of targets.}
    \label{fig:dr12_dr13_hr}
\end{figure*}

\begin{figure*}
    \centering
    \includegraphics[width=\textwidth]{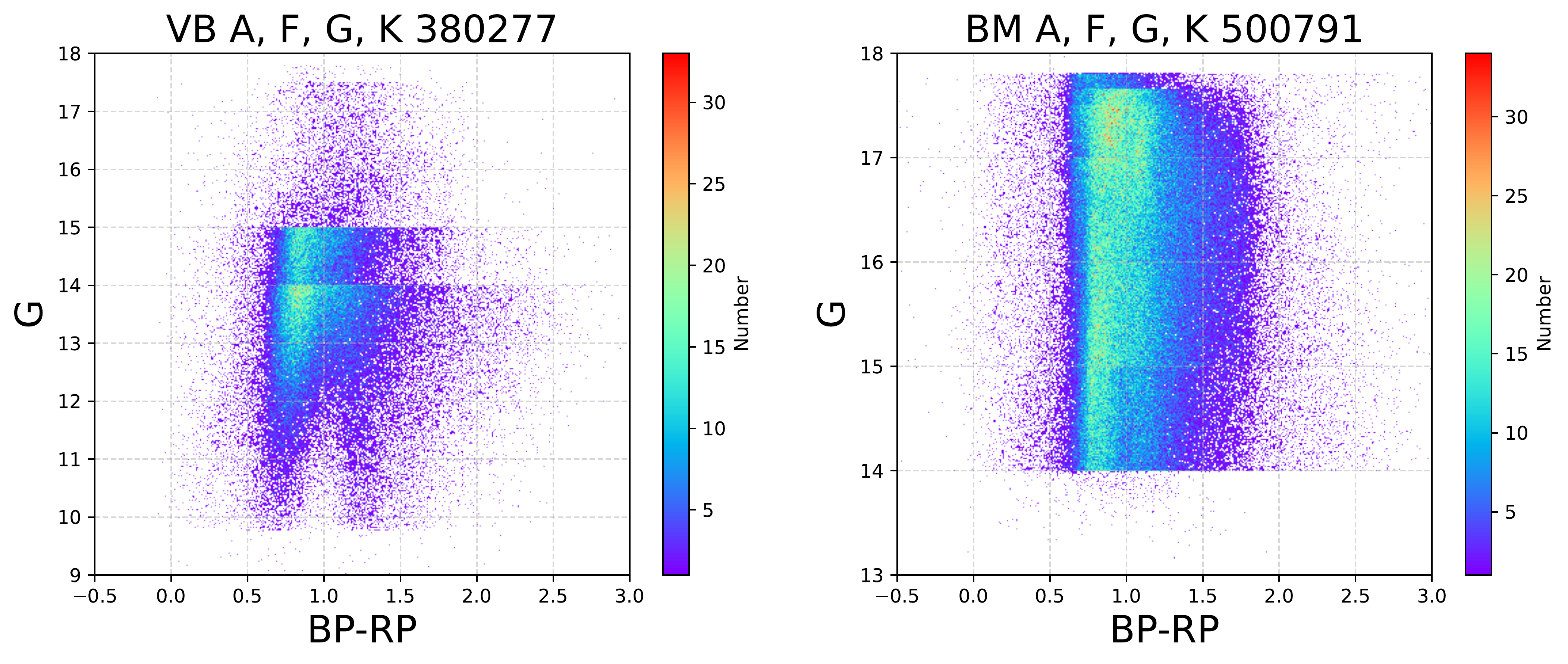}
    \caption{Color--magnitude distribution of A-, F-, G-, and K-type stars observed by LAMOST~III LRS in the VB (left) and BM (right) plates, color-coded by number density. The panel titles report the number of targets.}
    \label{fig:dr12_dr13_stellar_cmd}
\end{figure*}

\begin{figure*}
    \centering
    \includegraphics[width=\textwidth]{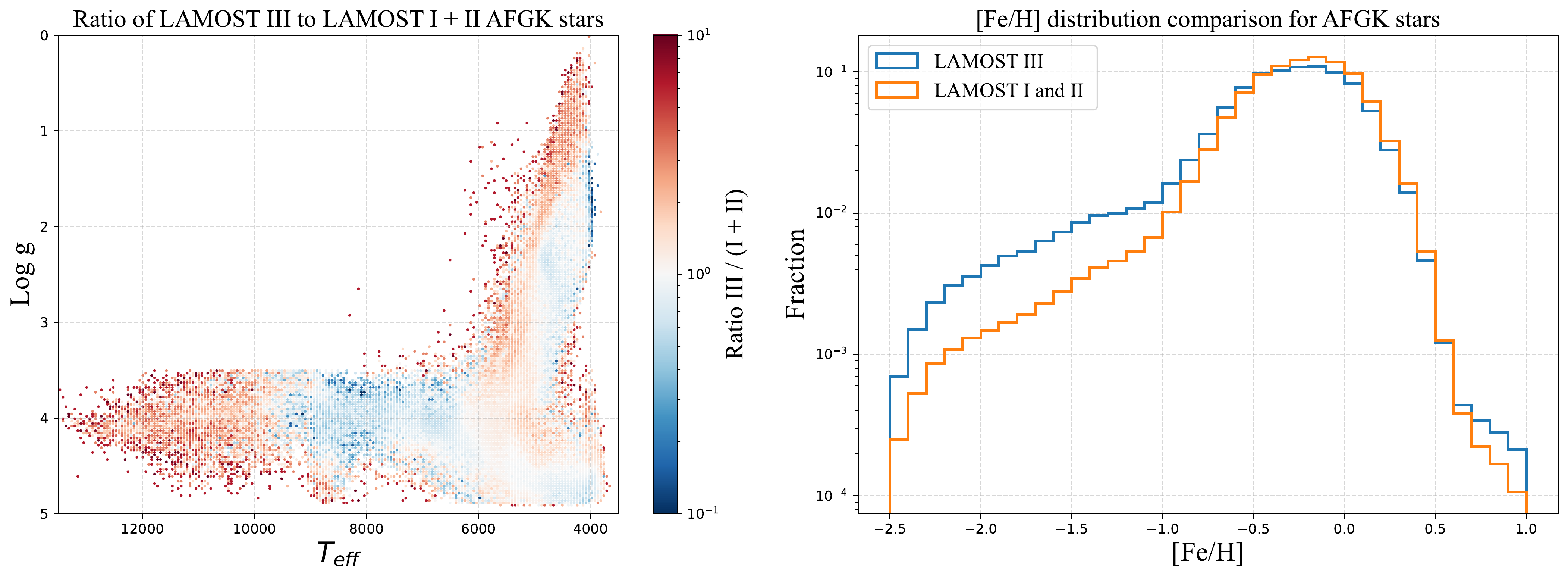}
    \caption{Comparison between LAMOST~III LRS and LAMOST~I+II in the HR diagram and [Fe/H] distribution. Left: ratio map in the HR diagram, where the color encodes (LAMOST~III/LAMOST~I+II) of the per-bin stellar fractions. Right: log-scaled [Fe/H] distributions for LAMOST~III LRS (blue) and LAMOST~I+II (orange).}
    \label{fig:dr12_dr13_feh}
\end{figure*}

As of June 2025, 
LAMOST~III LRS has observed 881,068 A-, F-, G-, and K-type unique stars from 1,065,500 spectra, with 380,277 and 500,791 unique stars obtained on the VB and BM plates, respectively.
Their distributions in the HR diagram and the $T_{\rm eff}$--[Fe/H] plane are shown in the top and bottom panels of Figure \ref{fig:dr12_dr13_hr}.
The distribution in the Gaia color-magnitude diagram is shown in Figure \ref{fig:dr12_dr13_stellar_cmd}.
To compare LAMOST~I+II and LAMOST~III LRS, we bin the HR diagram and compute, in each bin, the fraction of A-, F-, G-, and K-type unique stars relative to the full sample; the left panel of Figure \ref{fig:dr12_dr13_feh} shows the ratio of these fractions (LAMOST~III/LAMOST~I+II).
In this panel, red (blue) regions indicate bins where LAMOST~III LRS contributes a higher (lower) fraction than LAMOST~I+II.
As expected, LAMOST~III LRS preferentially samples O- and B-type stars, VMP stars, subgiants, turn-off stars, and other special targets.
The right panel of Figure \ref{fig:dr12_dr13_feh} shows the [Fe/H] distribution, indicating that the MP/VMP fractions relative to the AFGK star sample in LAMOST~III LRS are significantly higher than in LAMOST~I+II (9.0\% vs.~3.8\% and 1.1\% vs.~0.4\%, respectively).
LAMOST~III LRS contains 79,486 MP stars and 9,857 VMP unique stars.
Giants make up 23.05\%, 12.07\%, and 9.84\% of the observed LAMOST~III AFGK sample at low ($|b|<30^\circ$), intermediate ($30^\circ \le|b|\le60^\circ$), and high ($|b|>60^\circ$) Galactic latitudes, respectively. These values exceed the corresponding fractions in LAMOST~I+II by 3.22\%, 2.35\%, and 2.24\%. The overall fractions are lower than might be expected because only priority-4 and -5 targets at low Galactic latitudes have very high giant fractions (Figures \ref{fig:mian_survey_point_vb} and \ref{fig:mian_survey_point_bm}). Many add-on catalog targets are also concentrated at low Galactic latitudes and have higher priority than the main survey; as a result, priority-4/5 observations contribute only 24.3\%/15.5\% of the AFGK sample at low Galactic latitudes. However, among all observed giants, the fraction beyond 10~kpc has increased by a factor of three, from $\sim$1.5\% in LAMOST~I+II to 4.5\% in LAMOST~III.

\section{Summary and Future Perspective}\label{sec:summary}

As the LAMOST~III LRS enters its mid-term stage, this paper reviews its scientific goals, survey design and target-selection strategy, observing progress, and future prospects. The survey is tailored to increase spectroscopic coverage of distant tracers, rare stellar populations, and other high-value targets that were underrepresented in LAMOST~I/II, and is organized around four main goals: mapping the structure and evolution of the Galactic disk; reconstructing the assembly and early chemical-enrichment history of the Galactic halo; improving stellar-parameter calibration and enabling advances in stellar astrophysics; and supporting time-domain studies of binaries and other peculiar systems. To meet these goals, LAMOST~III LRS integrates a main survey with dedicated add-on catalogs and a time-domain component, together with an updated target-selection scheme to boost observing efficiency for key populations. We summarize the survey implementation, progress through June~2025, and the expected scientific return over the remainder of LAMOST~III LRS.

\paragraph{Survey design and target selection.}
The LAMOST~III LRS consists of the main survey, the add-on catalogs, and the TD survey. Targets are separated into VB ($9<G<14$) and BM ($14<G<17.8$) plates.
The main-survey input catalog is constructed from Gaia DR3, with stellar atmospheric parameters, extinction, and absolute magnitudes derived for sources with Gaia BP/RP spectra. These measurements are further used to estimate $M_G$, $E(B-V)$, and intrinsic $(BP-RP)_0$ for the broader Gaia sample down to $G=17.8$, providing the basis for the subsequent target selection.
Within the main survey, priorities are set by an HRD weight (inverse local density in the HR diagram), a $G$-band weight (inverse local density in the $G$-magnitude distribution), and an enhancement factor (assigned according to different regions of the HR diagram), resulting in priorities for $\sim$100 million stars. 
The add-on catalogs further boost the observing fractions of high-value targets (e.g., halo stars, young stars, VMP stars, M dwarfs, variables, binaries, and extragalactic sources), generally at higher priority than main-survey targets.
The TD survey repeatedly scans three eFEDS plates and ten low-Galactic-latitude plates, with an emphasis on binaries and other peculiar systems.

\paragraph{Current status (June 2025).}
LAMOST~III LRS has collected 1.53 million low-resolution spectra from 658 plates, delivering atmospheric parameters and radial velocities for 1,264,052 unique sources (456,776 on VB plates and 807,276 on BM plates).
For the VB/BM plates, 83.11\%/38.52\% of sources have $\mathrm{S/N}_g>20$, and 50.03\%/15.1\% have $\mathrm{S/N}_g>50$.
The current dataset contains 881,068 A-, F-, G-, and K-type stars and shows enhanced relative fractions of O/B stars, VMP stars, subgiants, turn-off stars, and other special targets compared to LAMOST~I+II.
By June 2025, 79,486 metal-poor (MP) stars and 9,857 very metal-poor (VMP) stars have been observed, corresponding to 37\% and 44\% of the MP and VMP samples from LAMOST~I+II, respectively.

\paragraph{Future perspective.}
The facility is undergoing a staged upgrade of the fiber system from 4,000 to a planned 8,000 fibers, with installation and testing expected to be completed by 2027.
The LAMOST~III LRS is expected to collect $\sim$3.5 million high-quality spectra from $\sim$600 VB and $\sim$1000 BM plates.
Key projected data products include $\sim$1.2 million giants (about 1.6 times earlier surveys, with distant giants expected to triple), enabling improved constraints on the halo's structure, chemistry, and kinematics and, with dynamical modeling, on the Galactic potential and mass distribution.
The survey is also expected to include $\sim$600,000 young stars (about four times previous yields) and to deliver reliable age estimates for $\sim$1.2 million stars, strengthening studies of disk evolution and the Milky Way's assembly history.
Finally, the identification of $\sim$20,000 VMP stars will enable efficient searches for second-generation stars and stringent tests of early chemical enrichment, nucleosynthesis, and supernova-yield models.
Beyond these core samples, LAMOST~III LRS will build specialized datasets (e.g., calibration stars; peculiar, variable, and time-domain samples; quasars; and extragalactic galaxies), positioning the survey to deliver broad impact across Galactic and stellar astrophysics, the interstellar medium, and data-driven astronomy.

\Acknowledgements{
This work is supported by the National Key R\&D Program of China via 2024YFA1611901 and 2024YFA1611601, and the National Natural Science Foundation of China through the projects 12222301, and 12173007. 
JNF acknowledges support from the China Manned Space Program with grant no. CMS-CSST-2025-A13, the Tianchi Talent Introduction Plan, and the Central Guidance for Local Science and Technology Development Fund under No. ZYYD2025QY27.
HLY acknowledges support from the Strategic Priority Research Program of Chinese Academy of Sciences under grant No. XDB1160101 and the National Natural Science Foundation of China through the project 12373036. 
LY acknowledges support from Beijing Natural Science Foundation (Grant No.1264072). 
This work made use of the data from LAMOST (Large Sky Area Multi-Object Fiber Spectroscopic Telescope, also known as the Guoshoujing Telescope) (\url{https://cstr.cn/31118.02.LAMOST}). LAMOST is a Chinese national mega-science facility, operated by National Astronomical Observatories, Chinese Academy of Sciences. This work has made use of data from the European Space Agency (ESA) mission {\it Gaia} (\url{https://www.cosmos.esa.int/gaia}), processed by the Gaia Data Processing and Analysis Consortium (DPAC, \url{https://www.cosmos.esa.int/web/gaia/dpac/ consortium}). Funding for the DPAC has been provided by national institutions, in particular, the institutions participating in the Gaia Multilateral Agreement.
}
\InterestConflict{}

\bibliographystyle{scpma}
\bibliography{lamost_three}

\appendix
\setcounter{figure}{0}
\renewcommand{\theHfigure}{A.\arabic{figure}}

\section{Validation of the Input Catalog for the Main Survey}\label{sec:appendix}

\begin{figure*}
    \includegraphics[width=1\linewidth]{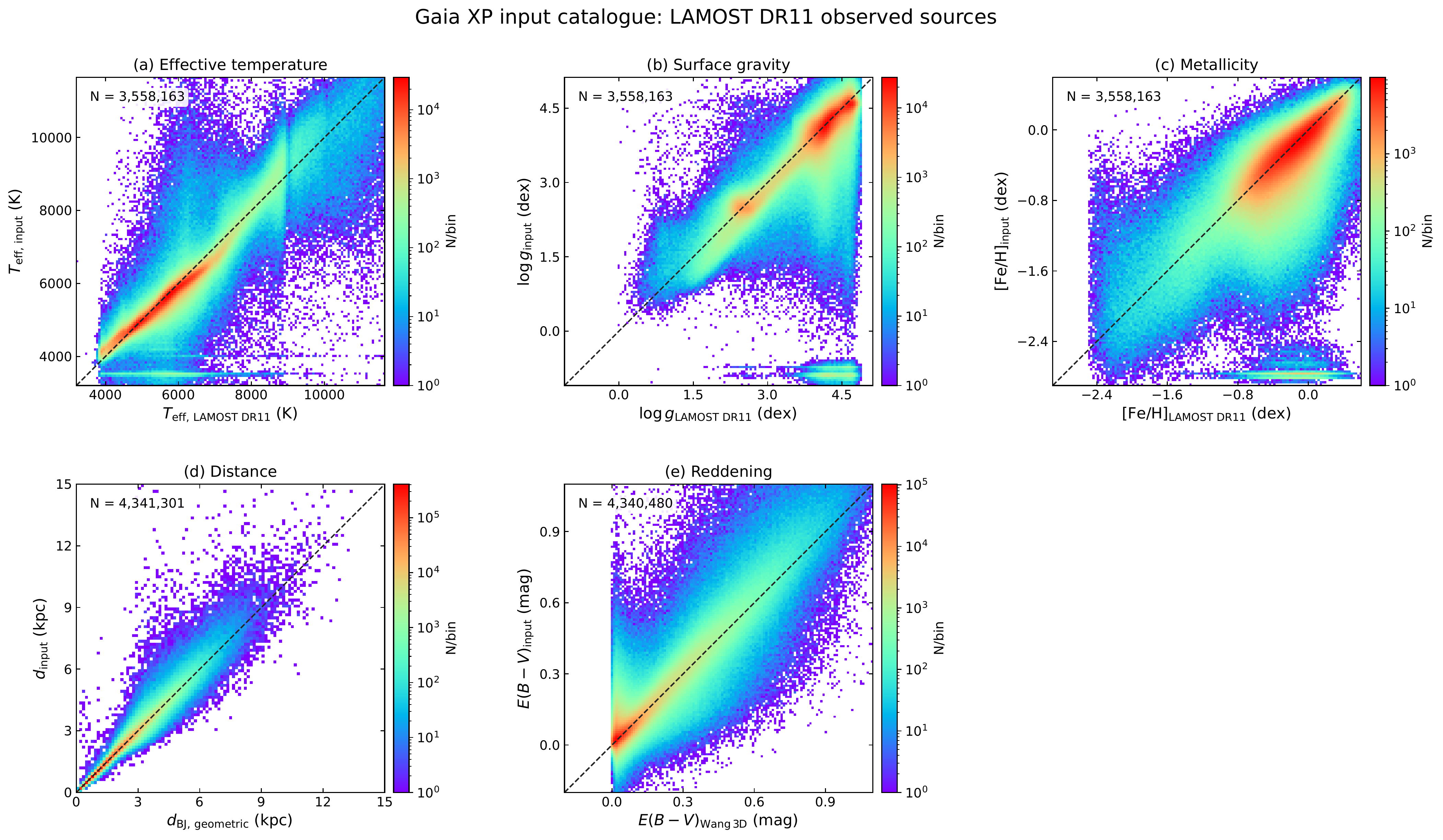}
    \caption{Comparison of stellar parameters in the Gaia XP input catalog with independent measurements. Panels (a)--(c) compare $T_{\rm eff}$, $\log g$, and [Fe/H] with LAMOST DR11, while panels (d) and (e) compare distances and reddening with Bailer-Jones et al.~\cite{BailerJones2021} and Wang et al.~\cite{Wang2025}, respectively. The number of sources included in each panel is indicated in the upper-left corner. The color scale represents the logarithmic number density, and the dashed lines indicate one-to-one relations.}
    \label{comparasion}
\end{figure*}

To assess the reliability of the stellar parameters in our input catalog, we cross-matched it against LAMOST DR11, the Gaia EDR3 distance catalog of Bailer-Jones et al.~\cite{BailerJones2021}, and the all-sky 3D dust map of Wang et al.~\cite{Wang2025}. 
For sources with multiple LAMOST observations, we retained the spectrum with the highest ${S/N}_g$.
The comparison of atmospheric parameters was restricted to sources with ${S/N}_g>20$, whereas the validation of distances and reddening was limited to sources with S/N in parallax larger than 5.
In panel (e) of Figure~\ref{comparasion}, we additionally excluded 821 sources because their distances exceed the distance coverage of the Wang et al.~\cite{Wang2025} dust map.
The comparisons are presented in Figure~\ref{comparasion}.

Overall, the parameters in our input catalog agree well with independent measurements.
The standard deviations ($\sigma$) of the differences in $T_{\rm eff}$, $\log g$, [Fe/H], and $E(B-V)$ are 140\,K, 0.16\,dex, 0.13\,dex, and 0.02\,mag, respectively.
The standard deviation of the relative distance difference is 1.15\%, 9.23\%, and 12.95\% for sources with Bailer-Jones et al.~\cite{BailerJones2021} distances $<5$~kpc, 5--10~kpc, and $>10$~kpc, respectively.
For sources within 5~kpc, both distance estimates are largely parallax-driven, leading to close agreement.
Nevertheless, in the bottom-right corner of panel (b) in Figure~\ref{comparasion}, a subset of sources are classified as giants in the input catalog, whereas their LAMOST spectroscopic parameters suggest otherwise.
This discrepancy arises because these objects lack parallax measurements; there are only 29{,}537 such sources, corresponding to 0.8\% of the full sample.
The dense bands near the lower edges of panels (a) and (c) in Figure~\ref{comparasion} are likewise dominated by these no-parallax sources.
In panel (b) of Figure~\ref{comparasion}, a fraction of sources classified as dwarfs by LAMOST are assigned input-catalog labels spanning a wide range from dwarfs to giants.
Further analysis suggests that these objects typically have parallax S/N $<5$ and comparatively bright Gaia-parallax-based absolute magnitudes.
We therefore consider the input-catalog classifications to be more reliable in this regime.

\end{multicols}
\end{document}